\documentclass[aps,prd,groupedaddress,twocolumn,floatfix,showpacs,tightenlines]{revtex4-1}
\usepackage{graphicx}
\usepackage{color}

\newcommand{\beq}{\begin{equation}}
\newcommand{\eeq}{\end{equation}}
\newcommand{\bea}{\begin{eqnarray}}
\newcommand{\eea}{\end{eqnarray}}
\newcommand{\bes}{\begin{subequations}}
\newcommand{\ees}{\end{subequations}}

\begin{document}
\title{Gravitational Wave Beacons}
\author{Carlos O. Lousto}
\author{James Healy} 
\affiliation{Center for Computational Relativity and Gravitation,\\
School of Mathematical Sciences,
Rochester Institute of Technology, 85 Lomb Memorial Drive, Rochester,
New York 14623}
\date{\today}

\begin{abstract}
We explore spinning, precessing, unequal mass
binary black holes to display the long term orbital angular
momentum, $\vec{L}$, flip dynamics. We study two prototypical cases
of binaries with mass ratios $q=1/7$ 
and $q=1/15$ and a misaligned spin of the large black hole (with
an intrinsic spin magnitude of $S_2/m_2^2=0.85$). 
We conduct full numerical simulations, for nearly 14 and 18 orbits respectively,
to evolve the binary down to merger and display a full $L$-flip cycle.
The pattern of radiation of such systems
is particularly interesting, displaying strong polarization-dependent 
variation of amplitudes at precessional frequencies, leading to
distinctive observational consequences for 
gravitational wave detectors.
These waveform features are strongly directional dependent and 
measurements of gravitational waves polarizations
can be exploited to disentangle the binary's 
parameters in various astrophysical scenarios. 
\end{abstract}
\pacs{04.25.dg, 04.25.Nx, 04.30.Db, 04.70.Bw}

\maketitle


\section{Introduction}\label{sec:intro}

The late orbital dynamics of spinning binary black holes remains a
fascinating area of research, especially since the 
numerical breakthroughs \cite{Pretorius:2005gq, Campanelli:2005dd, Baker:2005vv}
solved the binary black hole problem, making it possible to study
these systems via supercomputer simulations.
The understanding of the role individual spins play during the late binary dynamics
is particularly interesting since it provides the means to extract these
astrophysically crucial parameters from current and future 
gravitational waves observations.

Among the notable spin effects (without Newtonian analogs)
observed in supercomputer simulations are the 
hangup effect~\cite{Campanelli:2006uy,Healy:2018swt},
which prompts or delays the merger of binary black holes depending on
the sign of the spin-orbit coupling, $\vec{S}\cdot\vec{L}$,
the large
recoils of the final black hole remnant
\cite{Campanelli:2007ew}, reaching up to $5000km/s$ \cite{Lousto:2011kp};
the flip-flop of black hole spins in a binary, passing from
aligned to antialigned periods with respect to the orbital angular 
momentum \cite{Lousto:2015uwa};
and the alignment instability \cite{Kesden:2014sla}
(a case of imaginary flip-flop frequencies \cite{Lousto:2016nlp}).

In this paper we discuss a two case study of spinning binaries 
that lead to a total flip of the orbital angular momentum. These 
{\it beaconing} cases,
unlike the flip-flop of spins mentioned above coupling spin-spin,
may arise
during the transitional precession between the two single precession stages
(the initial $\vec{L}$-dominated and the final $\vec{S}$-dominated dynamics
\cite{Apostolatos94}) but applies to a much larger region of a binary's 
parameter space including antialigned spins.
For small mass ratio binary systems ($q=m_1/m_2<1/4$) in retrograde orbits (not
necessarily exactly antialigned) the beaconing stage occurs at 
a relatively common rate before merger.

We are interested in studying the case where this transition occurs
in the strong dynamical period of late inspiral and merger (as opposed
to the previously studied case of a low post-Newtonian order regime
\cite{Apostolatos94}) and display its potential observational consequences 
for different
gravitational wave detectors. 
For instance, the observation of orbiting stellar mass binary
black holes by LISA \cite{DelPozzo:2017kme} can not only inform earth based observatories
of this binary entering in band \cite{Tso:2018pdv}, but also, in turn, 
by determining the precession
parameters near merger, will lead to strong constraints on the precursor
binary observed in LISA band.

\section{Numerical Techniques}\label{sec:nr}

We evolve the binary black hole data sets using the {\sc
LazEv}~\cite{Zlochower:2005bj} implementation of the moving puncture
approach~\cite{Campanelli:2005dd} with the conformal
function $W=\sqrt{\chi}=\exp(-2\phi)$ suggested by
Ref.~\cite{Marronetti:2007wz}.  For the run presented here, we use
centered, sixth-order finite differencing in
space~\cite{Lousto:2007rj}, a fourth-order Runge Kutta time
integrator, and a 7th-order Kreiss-Oliger dissipation operator.
Our code uses the {\sc EinsteinToolkit}~\cite{Loffler:2011ay,
einsteintoolkit} / {\sc Cactus}~\cite{cactus_web} /
{\sc Carpet}~\cite{Schnetter-etal-03b}
infrastructure.  The {\sc Carpet} mesh refinement driver provides a
``moving boxes'' style of mesh refinement. In this approach, refined
grids of fixed size are arranged about the coordinate centers of both
holes.  The evolution code then moves these fine grids about the
computational domain by following the trajectories of the two
black holes.
We use {\sc AHFinderDirect}~\cite{Thornburg2003:AH-finding} to locate
apparent horizons.  We measure the magnitude of the horizon spin using
the {\it isolated horizon} (IH) algorithm detailed in
Ref.~\cite{Dreyer02a} and as implemented in Ref.~\cite{Campanelli:2006fy}.

In the tables below, we measure radiated energy,
linear and angular momentum, in terms of the radiative Weyl
Scalar $\psi_4$, using the formulas provided in
Refs.~\cite{Campanelli:1998jv,Lousto:2007mh}.
We extract the radiated 
energy-momentum at finite radius and extrapolate to $r=\infty$
with the
perturbative extrapolation described in Ref.~\cite{Nakano:2015pta}.
Quasicircular (low eccentricity) initial orbital parameters
are computed using the post-Newtonian techniques described in~\cite{Healy:2017zqj}. A varying $\eta(W)$ gauge parameter \cite{Lousto:2010ut} 
has been used for the $q=1/15$ evolutions.

Numerous convergence studies of our simulations have been performed.
In Appendix A of Ref.~\cite{Healy:2014yta}, we
performed a detailed error analysis of a prototypical configuration with
equal mass and spins aligned/antialigned with respect to the orbital
angular momentum. We varied the initial separation of the binary, the
resolutions, grid structure, waveform extraction radii, and the number
of $\ell$ modes used in the construction of the radiative quantities.
In Appendix B of Ref.~\cite{Healy:2016lce}, we calculated the
finite observer location errors and performed convergence studies
(with three different resolutions) for typical runs with mass ratio
$(q=1,3/4,1/2,1/3)$. Further convergence studies reaching down to
$q=1/10$ for nonspinning binaries are reported in Ref.~\cite{Healy:2017mvh}.
 For very highly spinning black holes ($s/m^2=0.99$)
convergence of evolutions was studied in Ref. \cite{Zlochower:2017bbg}
and for ($s/m^2=0.95$) in Ref. \cite{Healy:2017vuz}.
 A cross verification (including convergence of higher modes) with the totally
independent SpEC code has been performed in Refs. \cite{Lovelace:2016uwp,Healy:2017abq}
for GW150914 and GW170104 targeted simulations. A catalog of 320 RIT
simulations \cite{Healy:2017psd,Healy:2019jyf} can be found in 
\url{https://ccrg.rit.edu/~RITCatalog/}, but the region of parameter
space here discussed have not been previously studied with extensive 
convergence tests.
For our current simulations we monitor accuracy by measuring
the conservation of the individual horizon masses and spins during evolution,
as well as the level of satisfaction of the Hamiltonian and momentum constraints,
to ensure reaching an accuracy consistent with our main conclusions
(See Fig.~\ref{fig:mHaH}).

\section{Results}\label{sec:sim}

\begin{figure}
\centerline{
\includegraphics[width=0.75\columnwidth]{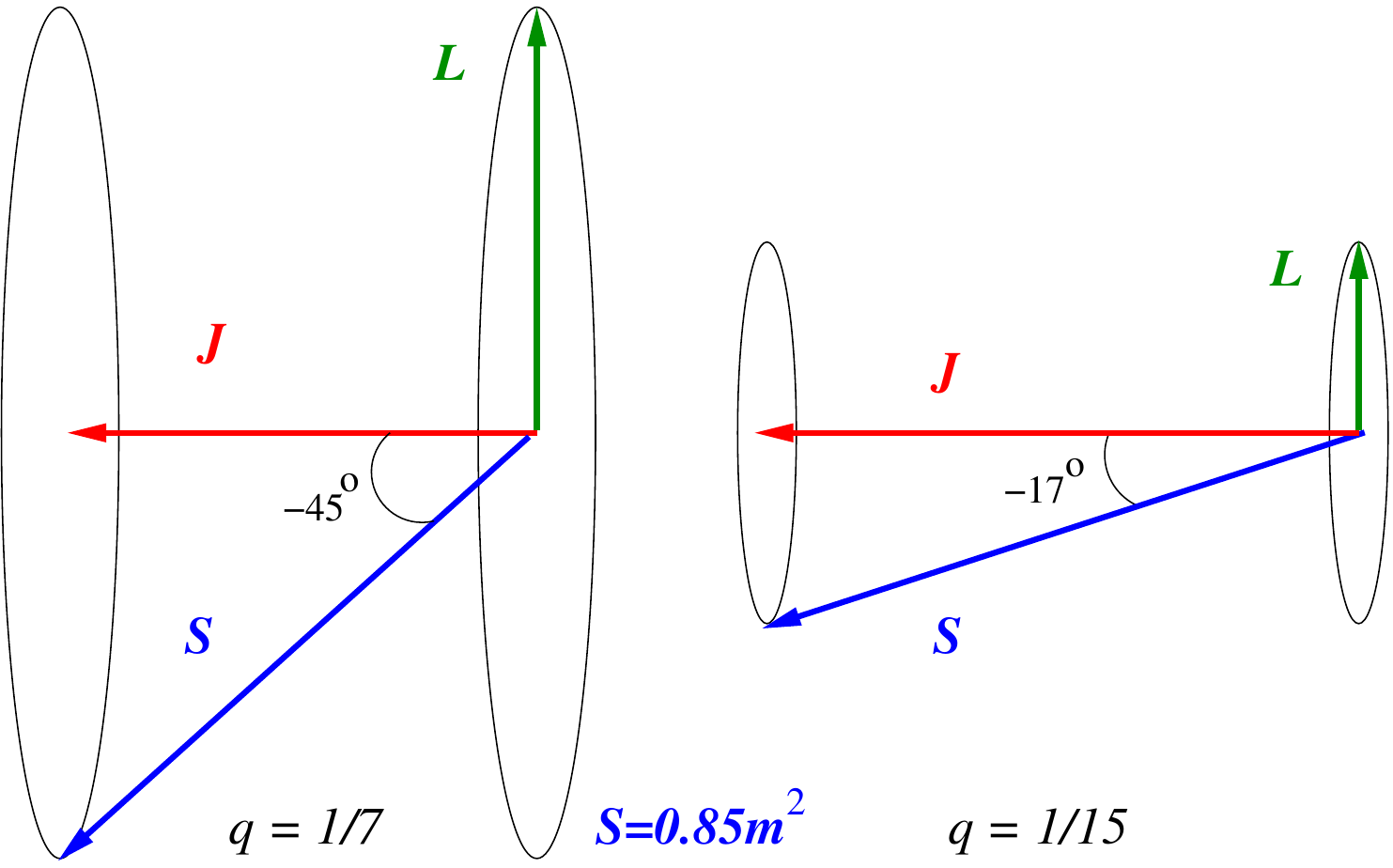}}
\caption{Initial configuration of the orbital angular momentum, $\vec{L}$,
  large hole spin, $\vec{S}$, and total momentum of the system, $\vec{J}$.
  Both the spin and the orbital angular momentum precess 
(counterclockwise) around $\vec{J}$ as the system evolves.
\label{fig:config}}
\end{figure}

In order to explore the L-flip precession regime during the
late inspiral and merger phase, we study two binary systems bearing
mass ratios of $q=1/7$ and $q=1/15$ respectively. Note that in a
previous study \cite{Lousto:2013vpa,Lousto:2013wta}, 
we estimated the need of $q<1/4$ in order for the
transitional precession to occur before merger. 
The chosen initial configurations are depicted
in Fig. \ref{fig:config} and given in more detail in Table \ref{tab:ID},
under the labels GWB7 and GWB15 respectively.
Two other reference set of runs with the same mass ratios but with 
nonprecessing and nonspinning black holes
are also reported here under the labels AS7, AS15, NS7, and NS15.

\begin{table*}
\caption{Initial data parameters for the quasi-circular
configurations with a smaller mass black hole (labeled as 1),
and a larger mass spinning black hole (labeled as 2). The punctures are located
at $\vec r_1 = (x_1,0,0)$ and $\vec r_2 = (x_2,0,0)$, have an initial simple
proper distance\cite{Lousto:2013oza} of $d$, with momenta
$P=\pm (P_r, P_t,0)$,  mass parameters
$m^p/M$, horizon (Christodoulou) masses $m^H/M$, total ADM mass
$M_{\rm ADM}$, the dimensionless spin of the
larger black hole $a_{2i}/m^H_2 = S_i/m_H^2$ (with $a_{2y}/m_2^H=0$ and $a^H_1/m_1^H=0$),
and the initial eccentricity, $e_0$.
The beaconing (nonprecessing,nonspinning) configurations are denoted by GWBX 
(ASX,NSX), where X gives the inverse mass ratio $m^H_2 / m^H_1$.
}\label{tab:ID}
\begin{ruledtabular}
\begin{tabular}{lccccccccccccc}
Run   & $x_1/M$ & $x_2/M$  & $d/M$ & $P_r/M$ & $P_t/M$ & $m^p_1/M$ & $m^p_2/M$ & $m^H_1/M$ & $m^H_2/M$ & $a_{2x}/m_2^H$ & $a_{2z}/m_2^H$ & $M_{\rm ADM}/M$ & $e_0$ \\
\hline
GWB7  & -11.594& 1.656 & 16.6 & -8.724e-5 & 0.03601 & 0.119940 & 0.459292 & 0.1250 & 0.8750 & -0.5993 & -0.6028 & 0.99657 & 0.0018 \\
GWB15 & -9.375 & 0.625 & 12.9 & -6.034e-5 & 0.02301 & 0.058896 & 0.492946 & 0.0625 & 0.9375 & -0.8105 & -0.2561 & 0.99766 & 0.0071 \\
AS7   & -11.594& 1.656 & 16.0 & -8.866e-5 & 0.03609 & 0.119934 & 0.720094 & 0.1250 & 0.8750 &  0.0000 & -0.6028 & 0.99657 & 0.0006 \\
AS15  &	-9.375 & 0.625 & 11.9 & -6.443e-5 & 0.02323 & 0.058879 & 0.908436 & 0.0625 & 0.9375 &  0.0000 & -0.2561	& 0.99770 & 0.0011 \\
NS7   &  -9.625& 1.375 & 13.1 & -1.425e-4 & 0.03964 & 0.118980 & 0.869837 & 0.1250 & 0.8750 &  0.0000 &  0.0000 & 0.99589 & 0.0006 \\
NS15  &  -9.375& 0.625 & 11.9 & -5.773e-5 & 0.02274 & 0.058913 & 0.934496 & 0.0625 & 0.8750 &  0.0000 &  0.0000 & 0.99764 & 0.0006 \\
\end{tabular}
\end{ruledtabular}
\end{table*}


For the sake of the simplicity of the analysis and its relatively
small influence on binaries of the chosen small mass ratios, we take the small hole to 
be spinless. We give a relatively large spin magnitude (0.85) to the larger
black hole in a configuration that initially corresponds to a polar orbit
of the small hole around the total angular momentum $\vec{J}$, i.e.
initially $\vec{L}$ and $\vec{J}$ are perpendicular to each other.

\begin{figure}
\centerline{
  \includegraphics[angle=270,width=0.49\columnwidth]{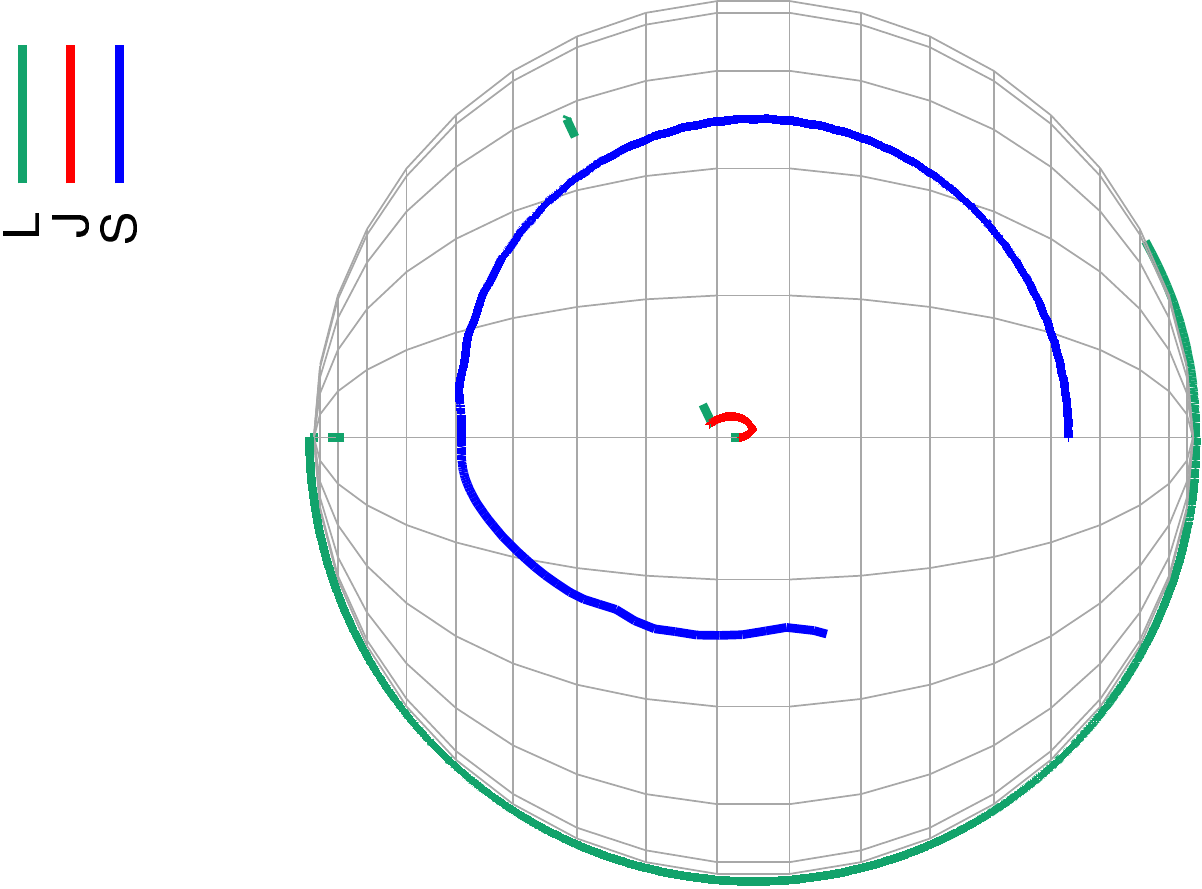}
    \includegraphics[angle=270,width=0.49\columnwidth]{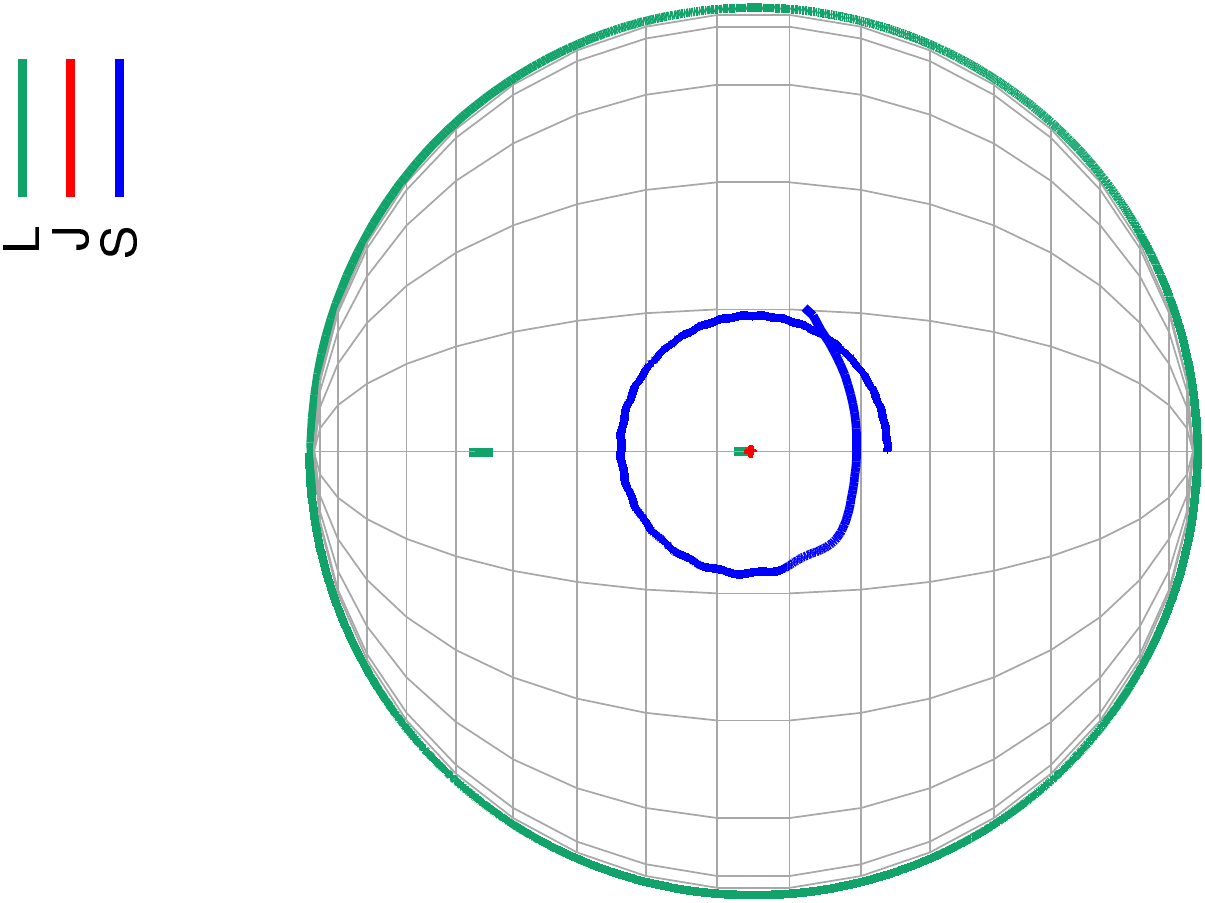}}
\caption{Evolution of the directions of the orbital angular momentum, $\vec{L}$,
  large hole spin, $\vec{S}$, and total momentum of the system, $\vec{J}$.
  The spin and the orbital angular momentum precess (counterclockwise)
  around $\vec{J}$. As
  the system evolves the loss of $\vec{L}$ does not seem to notably change
  the direction of $\vec{J}$ (See Table \ref{tab:remnant} below).
  Displayed on the left are the $q=1/7$ and on the right the $q=1/15$ binaries.
\label{fig:LSJ}}
\end{figure}

The subsequent evolution of the two configurations lead to 14.5 orbits before merger
arises for the $q=1/7$ case and 18 orbits before merger arises
for the $q=1/15$ case. The direction of the total momentum $\vec{J}$
shows a notable stability during the whole simulations as displayed
in Fig. \ref{fig:LSJ}, while its magnitude {\it grows} slightly due to a
tendency towards alignment of the total spin, roughly as 
$J\dot{J}=-L\dot{L}>0$. 
We also observe that the spin of the large hole, $\vec{S}_2$,
and the orbital angular momentum, $\vec{L}$, both precess 
counterclockwise around this almost
constant total momentum direction $\hat{J}=\vec{J}/J$. During the simulation
down to merger,
$\vec{L}$ completes 5/4 of a precession cycle (L-flip) around $\vec{J}$ for
the $q=1/15$ case and 3/4 of a precession cycle around $\vec{J}$ for
the $q=1/7$ case.


In order to qualitatively understand the basic dependence 
of the beaconing phenomena on the binary parameters,
we use a low order post-Newtonian analysis (See Eq. (3.2c) of \cite{Racine:2008qv})
with $\vec{S}_2\cdot\hat{L}=-\vec{L}\cdot\hat{L}$ initially,
to find a frequency of precession of $\vec{L}$
\beq\label{eq:OL}
M\Omega_L =  2 \alpha_2^J/(1+q)^2 (M/r)^3, 
\eeq
where $r$ is the coordinate separation of the holes, 
$\alpha_2^J=\vec{S}_2\cdot\hat{J}/m_2^2$ is the dimensionless spin of the large hole along $\vec{J}$ (perpendicular to $\vec{L}$),
$M=m_1+m_2$ the total mass of the system, and $q=m_1/m_2\leq1$ its mass ratio.

The critical separation radius, $r_c$, characterizing the middle of the transitional precession, where the condition $S_2^L=\vec{S}_2\cdot\hat{L}=-\vec{L}\cdot\hat{L}=-L$ is met (for this condition to exist, $r_c$ needs to be above the inclined isco radius \cite{Lousto:2009ka}) is hence
\beq\label{eq:rc}
(r_c/M)^{1/2}=(\alpha_2^L/2q)(1+\sqrt{1-8(q/\alpha_2^L)^2}).
\eeq

\begin{figure}
  \includegraphics[angle=270,width=\columnwidth]{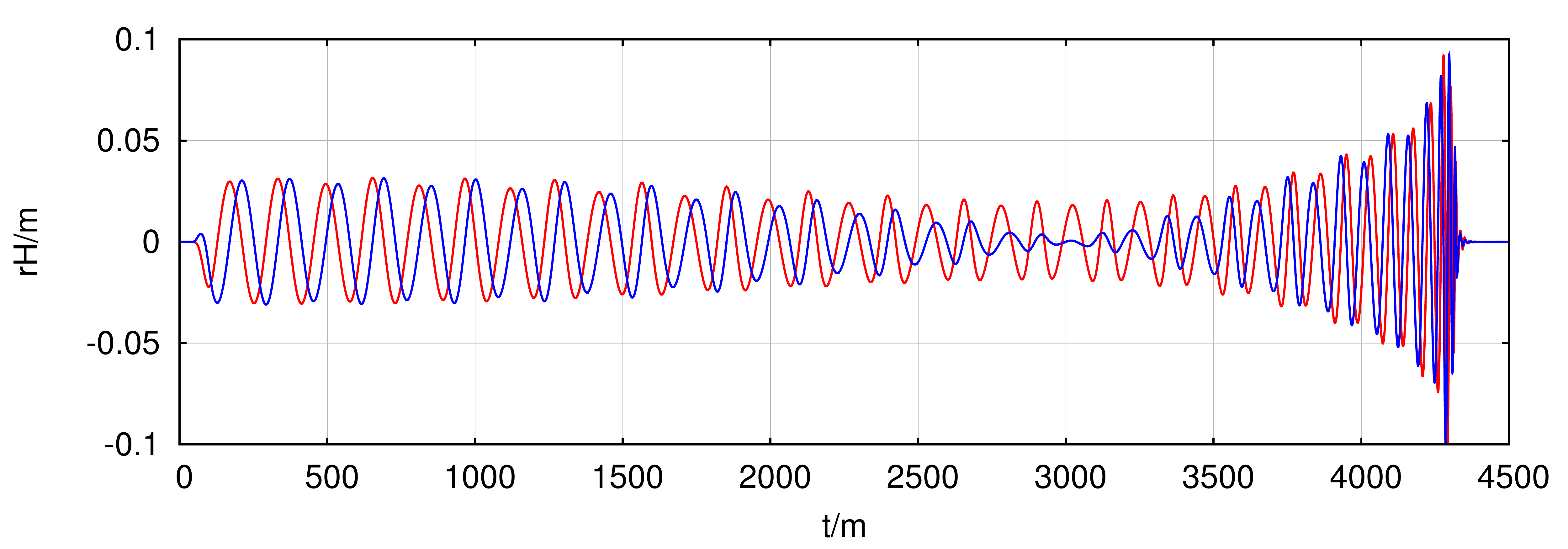}\\
  \includegraphics[angle=270,width=\columnwidth]{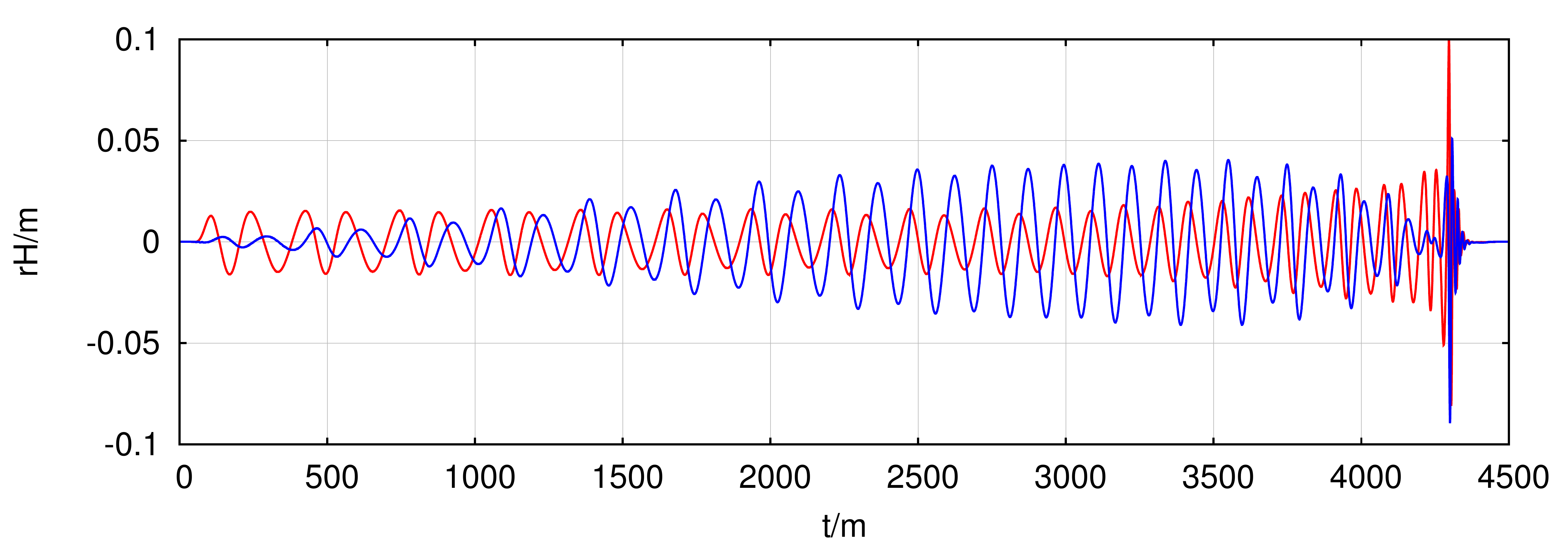}
  \caption{
    The two polarizations of the
    waveform strain of the system with mass ratio $q=1/7$ as seen
    from the z-axis (the initial direction of the orbital angular momentum) (above),
    and the same waveform strain as seen from the y-axis (below) reconstructed
    using modes up to $l_{max}=5$.
\label{fig:q7wvf}}
\end{figure}

For the initial separation of the $q=1/7$ simulation at ($r_c\approx13M$),
the formula
Eq.~(\ref{eq:OL}) leads to a period of $15042M$. While directly from the numerical
waveform displayed in Fig.~\ref{fig:q7wvf}
we obtain $3025M$, counting from $t=0$ to the first flip. 
So the estimated full beaconing period would at least be $4*3025 = 12100M$. 

A basic comparison of those values with the orbital frequency
in a PN expansion~\cite{Racine:2008kj}
\bea\label{eq:OO}
M\Omega_{orb}&&\approx\left(\frac{M}{r}\right)^{3/2}
-\frac{3+5q+3q^2}{2(1+q)^2}\left(\frac{M}{r}\right)^{5/2} 
\cr &&
-\frac{2\alpha_2^L(1+4q)+q\alpha_1^L(3+2q+5q^2)}{2(1+q)^3}\left(\frac{M}{r}\right)^{3},
\eea
leads to a period of $326M,$ at this initial configuration.


\begin{figure}
  \includegraphics[angle=270,width=\columnwidth]{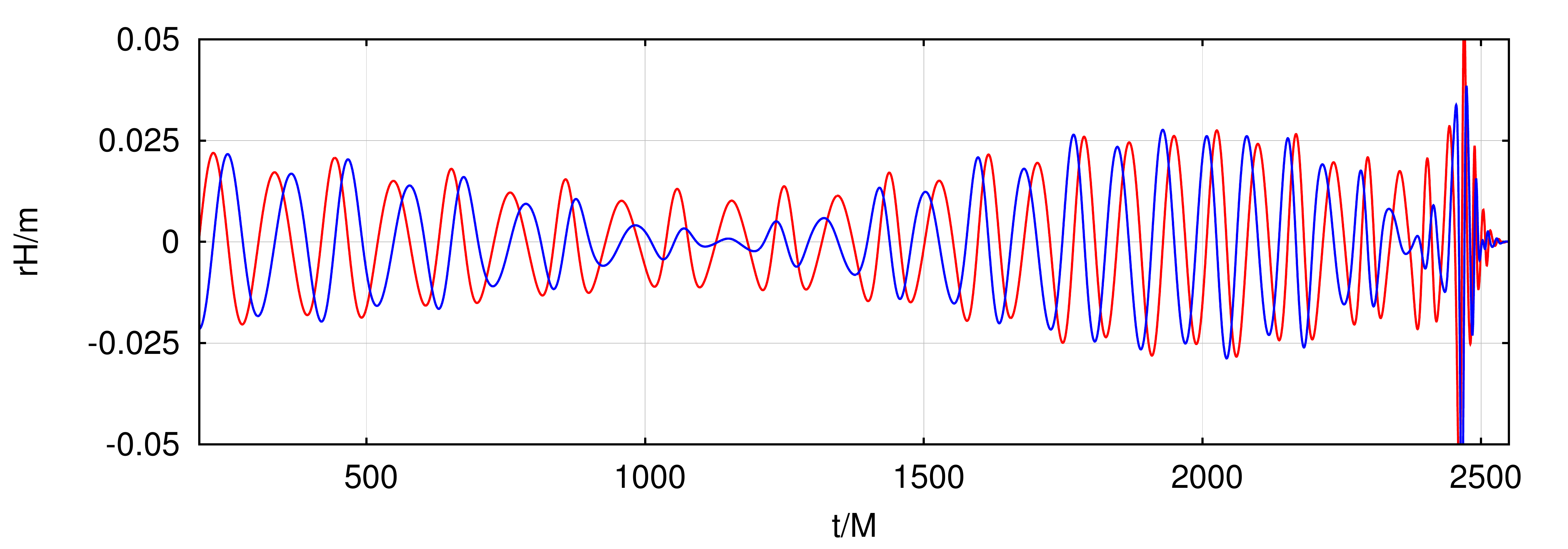}\\
  \includegraphics[angle=270,width=\columnwidth]{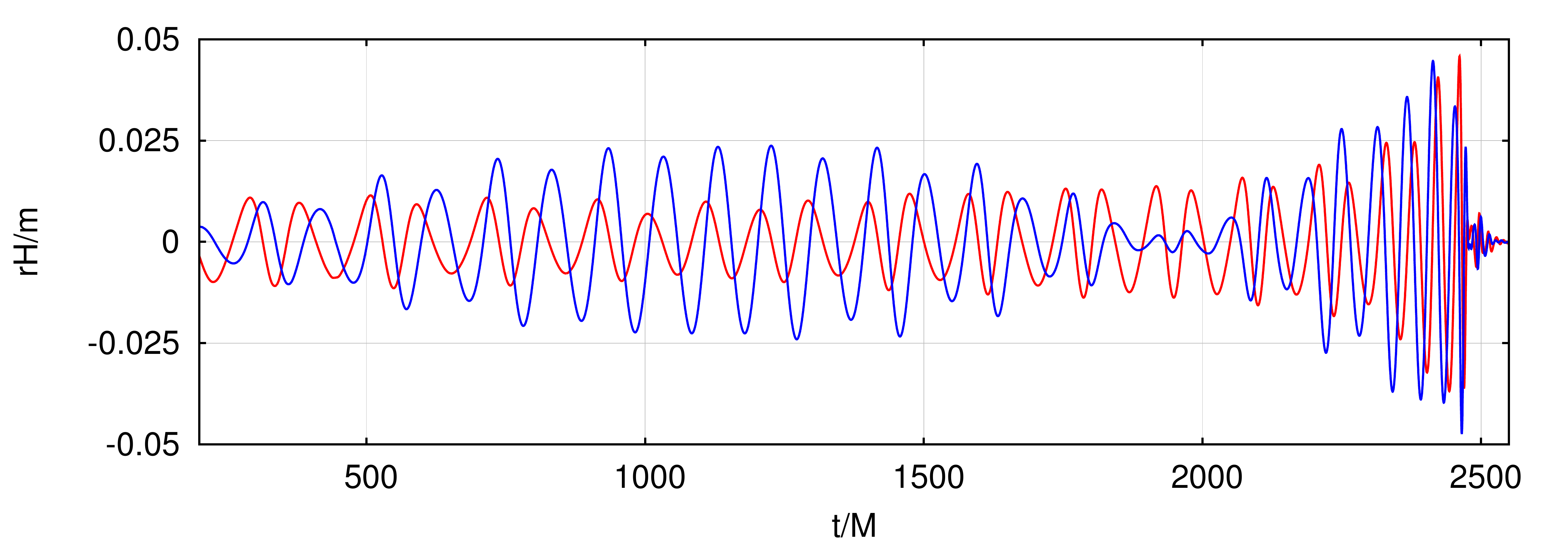}
  \caption{
    The two polarizations of the
waveform strain of the system with mass ratio $q=1/15$ as seen
    from the z-axis (the initial direction of the orbital angular momentum) (above),
    and the same waveform strain as seen from the y-axis (below) reconstructed
    using modes up to $l_{max}=5$.
\label{fig:q15wvf}}
\end{figure}

In the case of the simulation with mass ratio $q=1/15$
(at initial $r_c\approx10M$), guided by the waveform given
in Fig.~\ref{fig:q15wvf}, we observe that from
$t=0$ to peak 1 in the beaconing amplitude we get $\approx1111M$, from
peak 1 of the phase to zero-crossing $\approx948M$, from
zero-crossing to peak 2 $\approx280M$, and from
peak 2 to merger $\approx87M$. Using the initial first value, we
find a period of modulation of $\approx4444M$, which is in good
agreement with the period derived from Eq.~(\ref{eq:OL})
of $4410M$.  Similarly, the initial orbital period from Eq.~(\ref{eq:OO}) is $230M$.

\begin{figure}
  \includegraphics[angle=270,width=\columnwidth]{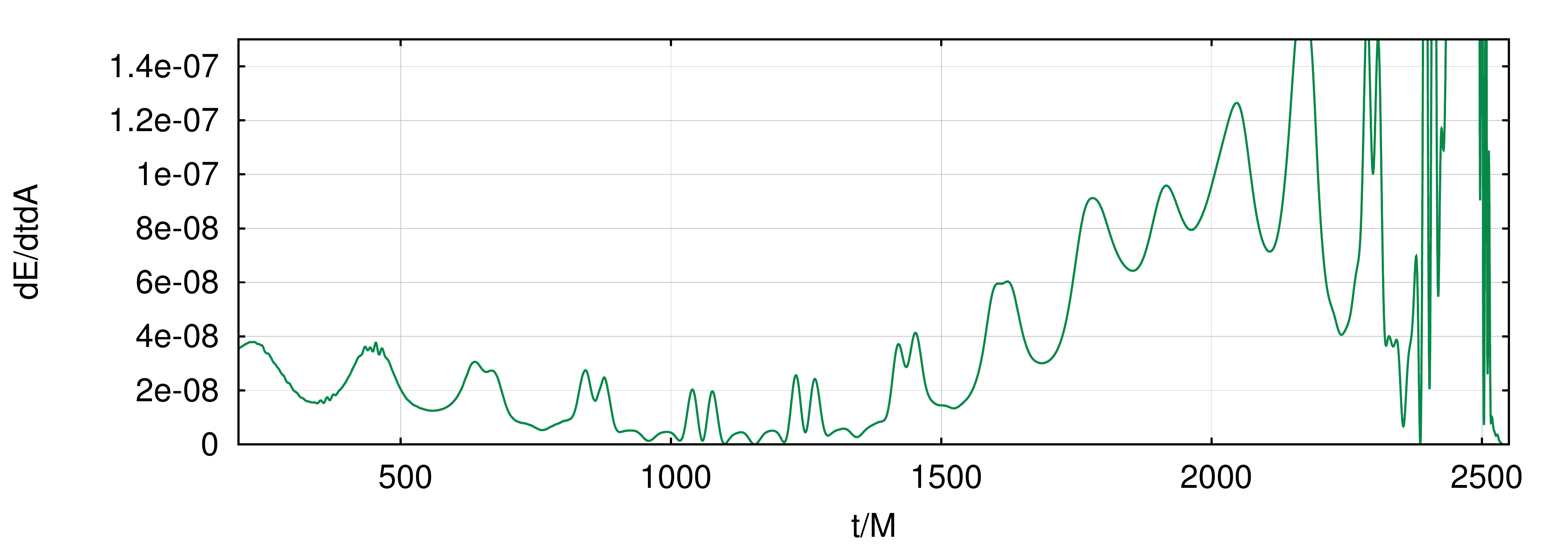}\\
  \includegraphics[angle=270,width=0.494\columnwidth]{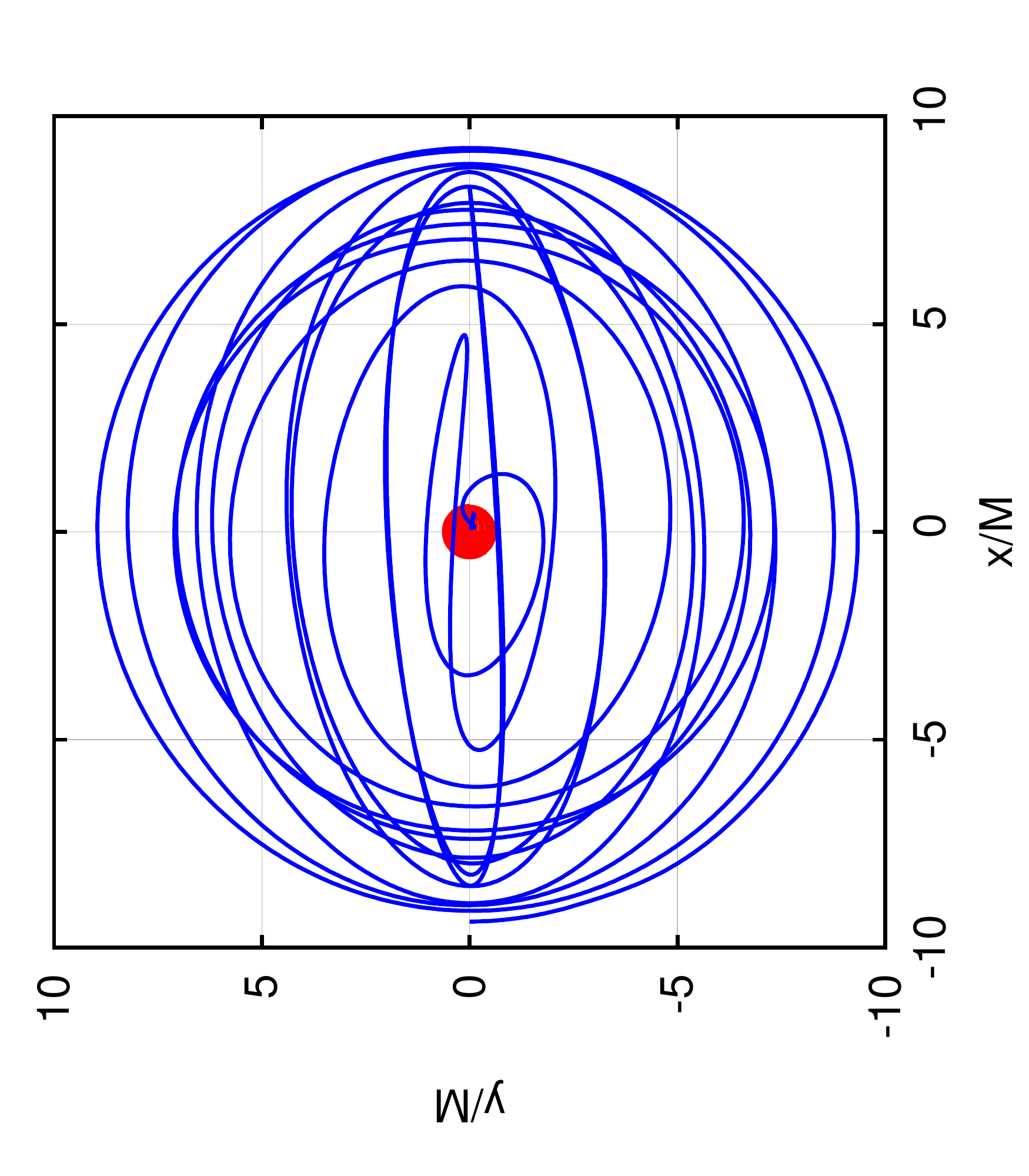}
  \includegraphics[angle=270,width=0.494\columnwidth]{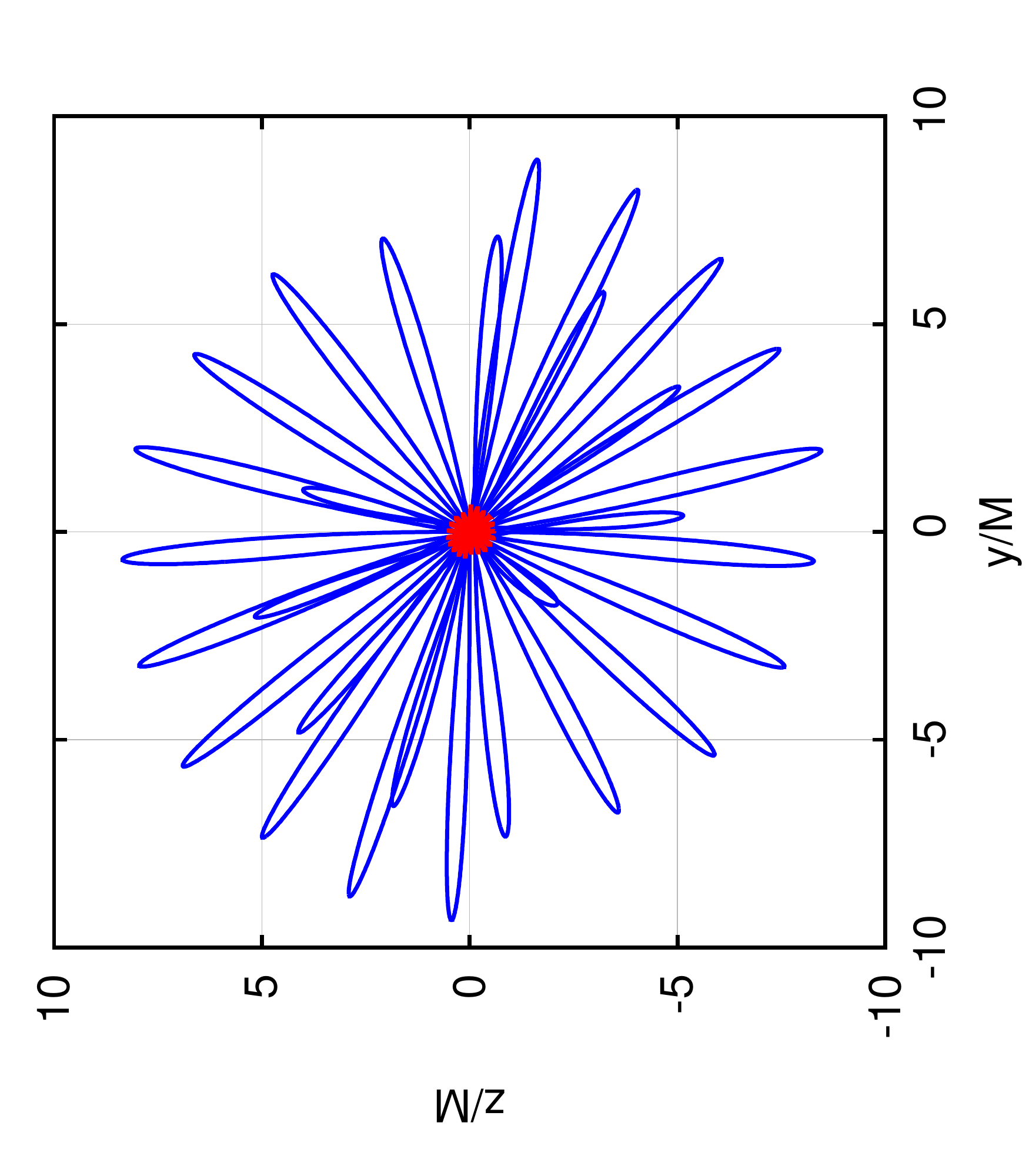}
  \caption{The beaconing effect displayed by the power radiated for the 
binary case with mass ratio $q=1/15$ as seen
    from the z-axis (the initial direction of the orbital angular momentum) (above),
    and (below) the detail of the black holes trajectories in the initial orbital plane (left) and seen from an observer along the x-axis (right).
\label{fig:q15beacon}}
\end{figure}

The complete flip of the orbital angular momentum has dramatic effects
on the pattern of gravitational radiation as seen by
observers far from the binary system. At directions near
perpendicular to the total angular momentum $\vec{J}$, radiation passes from
periods of full amplitude, when the system is face-on to
periods of strong suppression when the system is face-off.
This {\it beaconing} phenomena
is displayed in Fig.~\ref{fig:q15beacon} for the $q=1/15$ case,
and may have important
observational consequences regarding the parameter estimation of these systems
via measurements of the (two polarization of the) gravitational waves
(see below).

Additionally, the effect is strongly dependent on which polarization of
the gravitational waves we are observing. Figs.~\ref{fig:q7wvf} and \ref{fig:q15wvf} show
that while the amplitude of one of the polarizations remains slowly varying,
the other has dramatic variations (like a breathing mode) on the scale of this 
$\vec{L}$-precessional period. The black hole trajectories also display
a very rich pattern of polar precession until very late in the inspiral
motion (bottom panels of  Fig.~\ref{fig:q15beacon}.)

We provide in Table \ref{tab:remnant}
the remnant properties of the final black of the two simulations. 
The modeling of the final mass, spin and recoil of the final
merged black hole has been the subject of many studies
(See Ref.~\cite{Healy:2018swt} and references therein).
Also provided are the peak luminosity, amplitude and frequency,
that can be used for extending the modeling of those 
quantities \cite{Healy:2016lce,Jimenez-Forteza:2016oae}
to precessing binaries
with application to observations of gravitational waves from
black hole binaries.
Precessing binaries, as studied here, have very rich dynamics that
make it difficult to accurately model \cite{Lousto:2013wta,Zlochower:2015wga}.
Here we provide two useful new data points for forthcoming modeling
and testing.

\begin{table}
  \caption{Remnant properties of the beaconing simulations: final mass $M_{f}/M$ and spin $\alpha_{f}$,
    peak radiation $\mathcal{L}^{peak}$ and frequency $M\Omega^{peak}_{22}$,
    and recoil velocity $V_{recoil}$.
The final mass and spin are measured from the apparent horizon, and the
recoil velocity, peak luminosity, frequency, and amplitude are calculated from the gravitational waveforms.  Also given is the deviation of the angle between
the initial total angular momentum and final spin, $\Delta\theta$. 
\label{tab:remnant}}
\begin{ruledtabular}
\begin{tabular}{ccc}
  Case
&  $q=1/7$
&  $q=1/15$\\
\hline
$M_{f}/M$      & 0.989542 & 0.994071\\
$|\alpha_{f}|$ & 0.466308 & 0.716323\\
($\alpha_{f}^x$,$\alpha_{f}^y$,$\alpha_{f}^z)$ & (-0.4621,-0.0358,0.0510) & (-0.7162,0.0045,-0.0102)\\
$\Delta \theta$ & $6.19^\circ$ & $1.27^\circ$ \\
$V_{recoil}[km/s]$         &  275.2       & 78.1 \\
$\mathcal{L}^{peak}$       &  $1.4598\times10^{-4}$ & $5.4498\times10^{-5}$\\
$M\Omega^{peak}_{22}$      & -0.2975     & 0.2184\\
$\left(\frac{r}{M}\right)H^{peak}_{22}$ &  0.1688     & 0.0764
\end{tabular}
\end{ruledtabular}
\end{table}

\begin{figure}
  \includegraphics[angle=270,width=0.49\columnwidth]{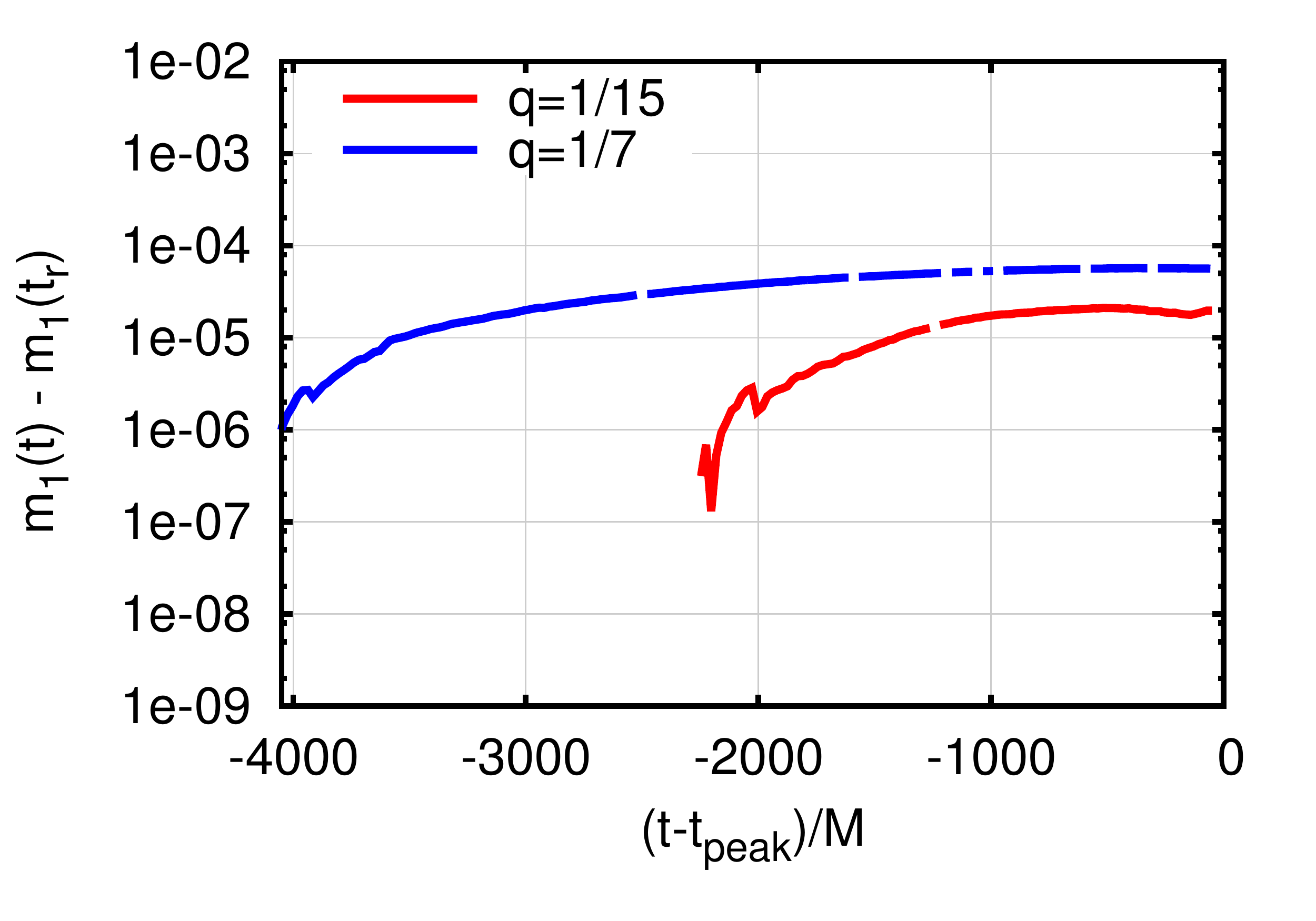}
  \includegraphics[angle=270,width=0.49\columnwidth]{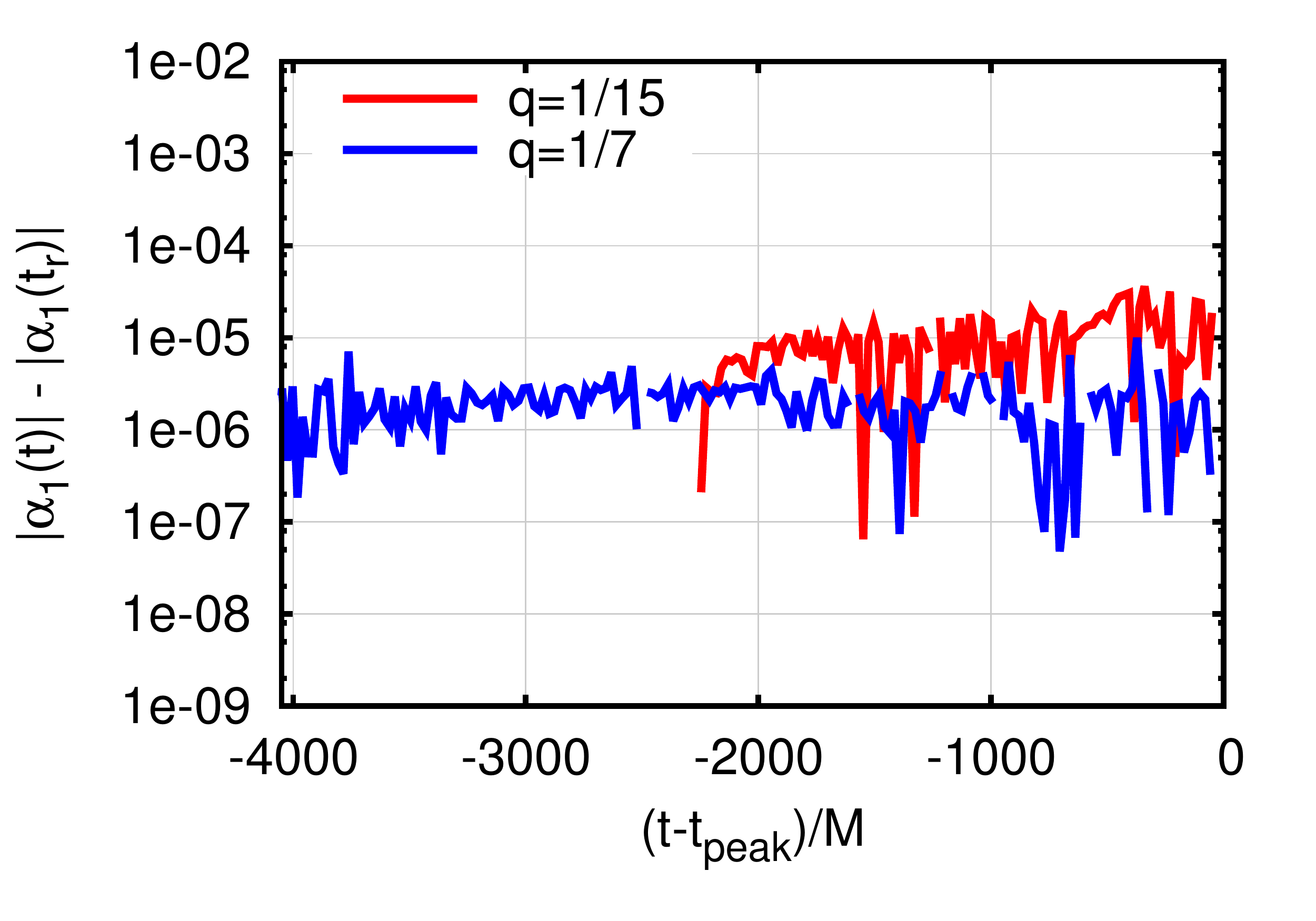}\\
  \includegraphics[angle=270,width=0.49\columnwidth]{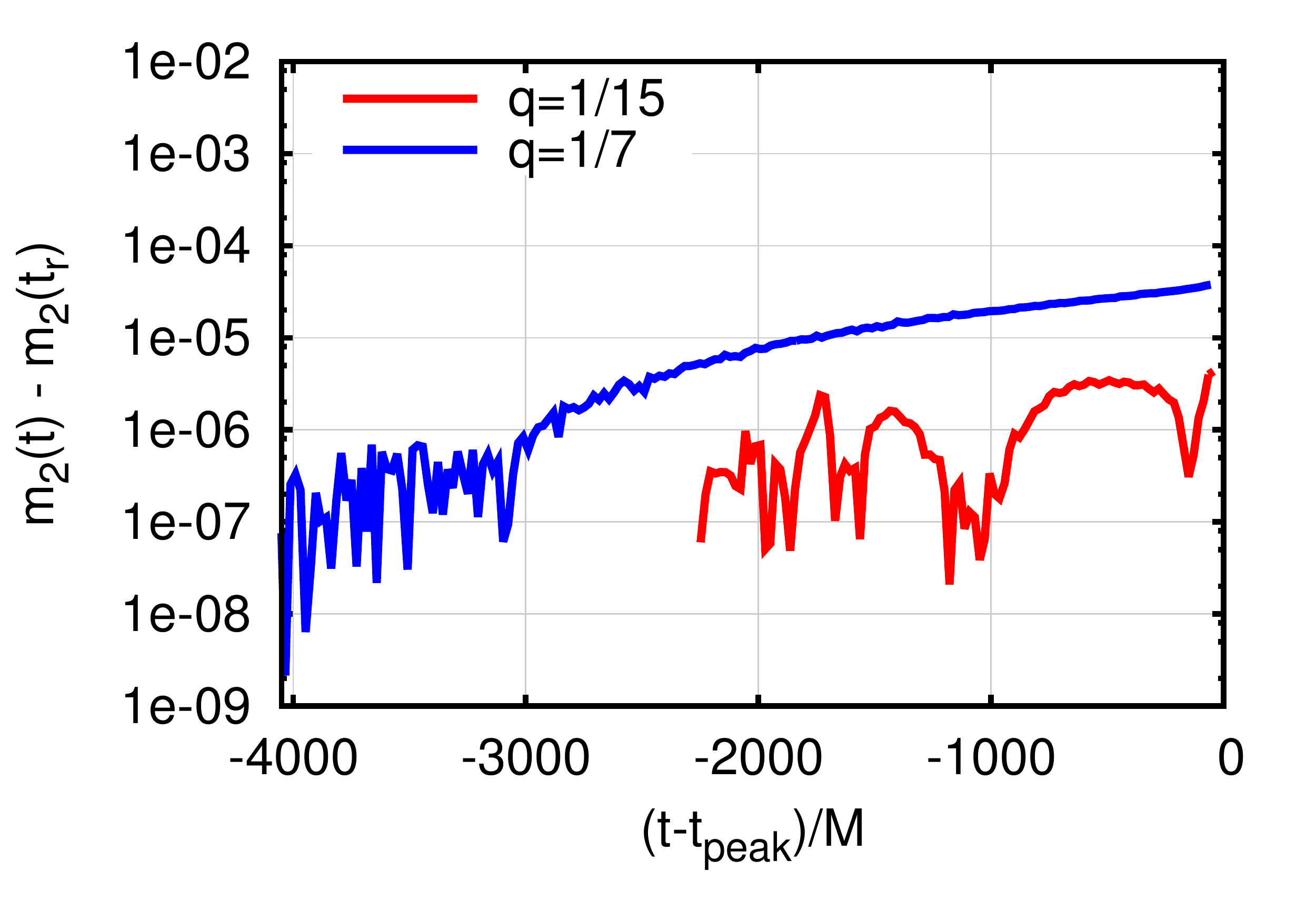}
  \includegraphics[angle=270,width=0.49\columnwidth]{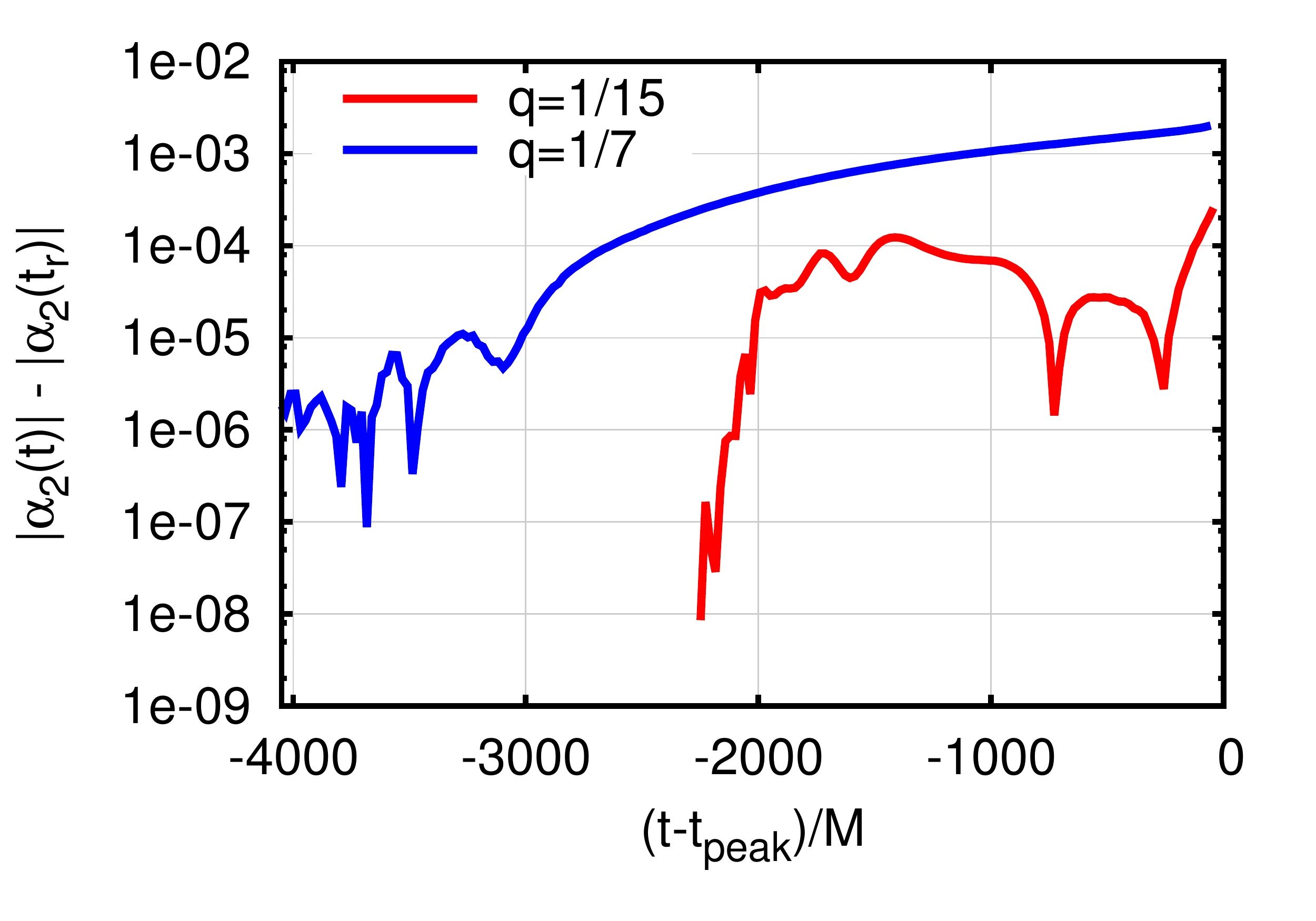}\\
  \includegraphics[angle=270,width=0.49\columnwidth]{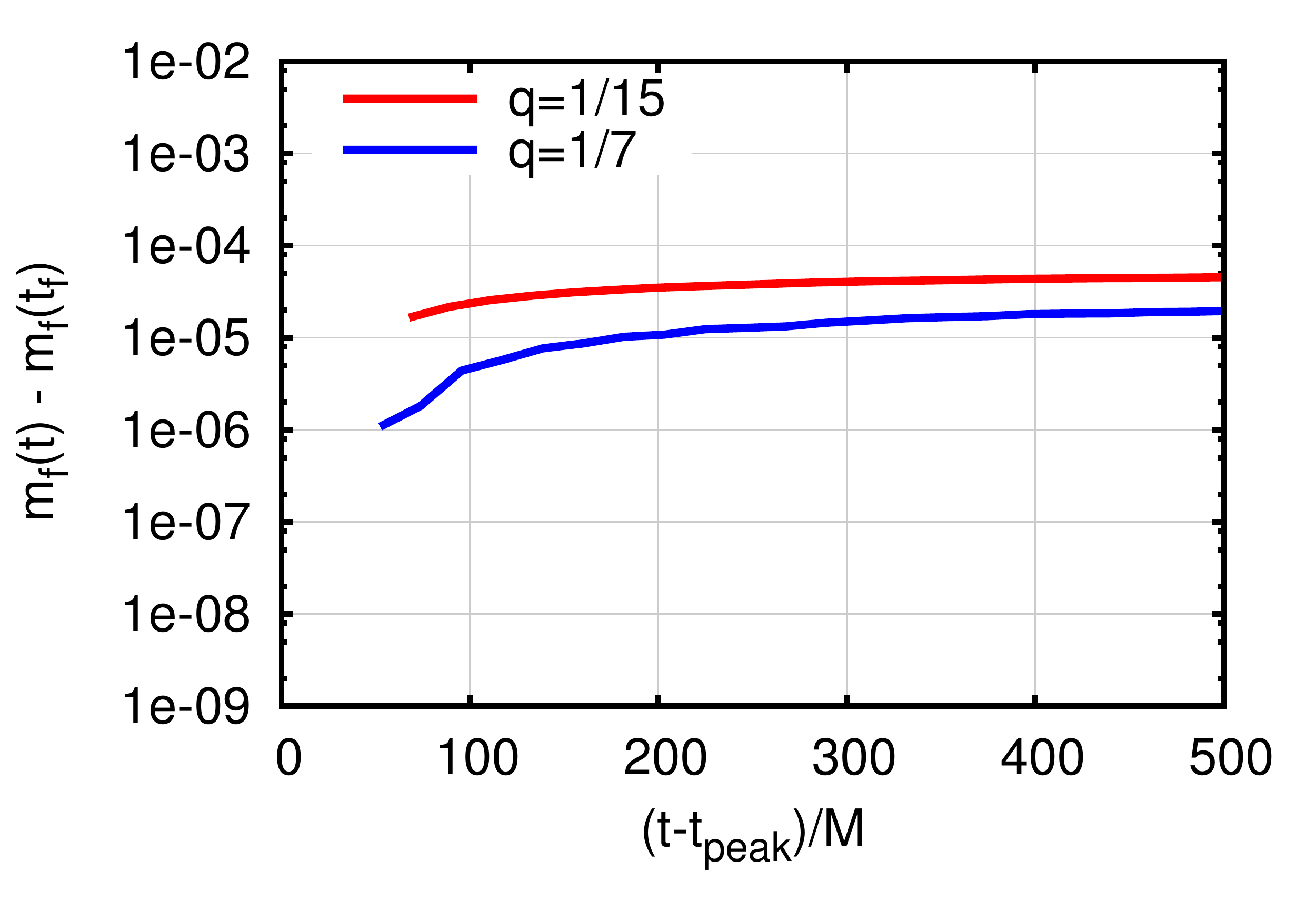}
  \includegraphics[angle=270,width=0.49\columnwidth]{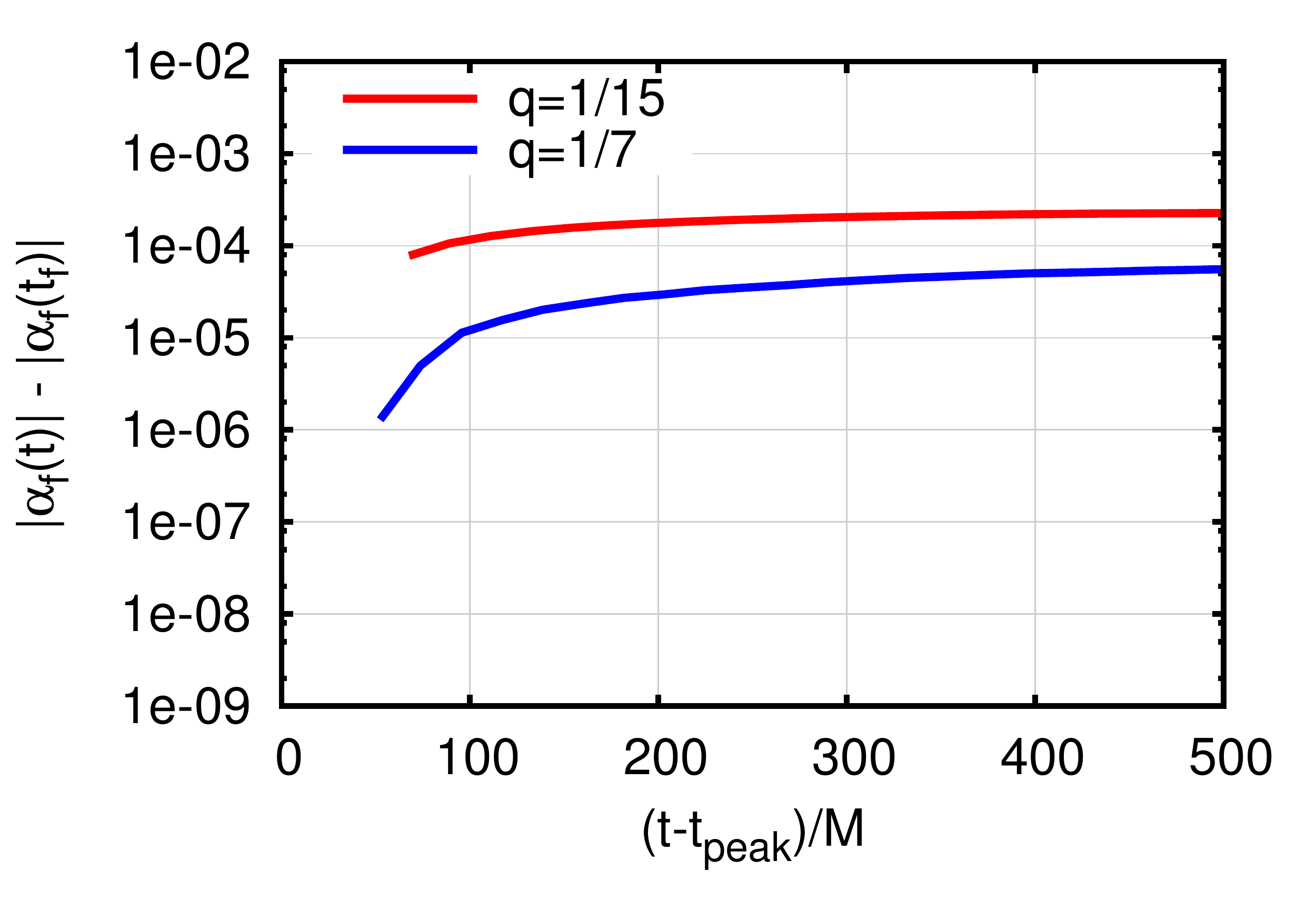}
  \caption{The evolution of the horizon masses (left) and spins (right) 
for the individual
black holes (labeled as 1 and 2) and the final merged hole (labeled as 3)
for both cases with mass ratios $q=1/7$ and $q=1/15$. All differences 
(for horizons 1 and 2)
are compared to the values at settling time $t_r = t_0 + 200M$ and to 50M after the
peak of the gravitational waves radiation $t_f = t_{peak} + 50M$ for horizon 3, 
where $t_0$ is the
start of the simulation. Since we are comparing both $q=1/7$ and $q=1/15$ runs
in the same plot and they have different run times, we shift the time of peak amplitude, $t_{peak}$, to 0 for a better display.
\label{fig:mHaH}}
\end{figure}

To provide a measure of the accuracy of the simulations we track the
individual horizons of the large and small black holes and compute
its masses and spins and display them in Fig.~\ref{fig:mHaH}. 
Those measurements are seen to be preserved at least to one 
part in $10^{4}$ in the cases of the masses and one part in $10^3$ in the cases of the spins.
Absorption of gravitational waves only plays an important role early
in the numerical evolution due to the initial data 
radiation content and we hence use as the reference measure a settling time of
$t_r=200M$. The merger of the two holes creates a larger final hole
with the mass and spin measured from its horizon as
reported in Table \ref{tab:remnant}.

\subsection{Comparisons with post-Newtonian waveforms and nonprecessing simulations}

We looked at several cases comparing the NR waveforms 
to the corresponding PN approximated waveform.  
For nonprecessing configurations, to calculate the amplitude of the 22 mode of the strain, 
we use 3.5PN orbital terms \cite{Blanchet:2008je,Faye:2012we}, 
and 1.5PN precessing spin terms \cite{Arun:2008kb}.  This calculation works for small 
inclination angles as well, but our beaconing cases in no way satisfy this approximation.  
This is evident when looking at the generated waveform.  While it does show a beaconing effect,
the amplitude is both qualitatively and quantitatively different, with both polarizations
losing power during the beaconing (compared to the NR case where this is only exhibited in
the imaginary part.)  For comparisons with the beaconing cases, we use the 1.5PN strain amplitude for generic inclination angles given in Appendix B of Ref.~\cite{Arun:2008kb}.
For the phasing information, in both cases we use a 3.5PN orbital evolution code with leading order
spin-spin terms and next-to-leading order spin-orbit terms~\cite{Buonanno:2005xu,Damour:2007nc}.
We can also use the NR trajectory information \cite{Campanelli:2008nk} for an even tighter hybrid comparison.

To add a further level of comparison we produced two new sets of full numerical simulations
with the same mass ratios $q=1/7$ and $q=1/15$ but for nonprecessing
and nonspinning binaries.
To calculate the match between NR and PN waveforms, we use a complex match 
\cite{Cho:2012ed} which factors in both polarizations of the waveforms,
a starting frequency of 30Hz, and the "ZERO\_DET\_high\_P" analytic 
noise curve for advanced LIGO \footnote{https://dcc.ligo.org/LIGO-T0900288/public}.
We choose the mass of the systems such that the waveforms start at 30Hz in LIGO band.
The results of the matches between the 22 mode of the strain for the NR
waveform and the PN waveform are given in Table~\ref{tab:matches} for the
6 different cases (nonprecessing, nonspinning and beaconing for $q=7$ and $q=15$.)  We show
the results for both PN waveforms where the phasing information in the amplitude
is supplied by PN or by NR. Since the complex match allows for maximization over
time and phase shift, we find an exceptionally good result when using the NR
phase information.  

\begin{table}
  \caption{Matches between PN and NR waveforms using the PN phase ($\phi_{PN}$) or
the NR phase ($\phi_{NR}$) for the PN waveform and the minimal SNR needed to distinguish
between the two waveforms, given the mismatch
($\mathrm{SNR}^2 \ge \frac{1}{1-\mathcal{M}}$.)  
\label{tab:matches}}
\begin{ruledtabular}
\begin{tabular}{lcccc}
Case      & $\mathcal{M}$ $\phi_{PN}$ & SNR $\phi_{PN}$ &  $\mathcal{M}$ $\phi_{NR}$ & SNR $\phi_{NR}$ \\
\hline
NS7     & 0.917 & 3.5 & 0.998 & 22.4 \\
NS15    & 0.693 & 1.8 & 0.994 & 12.9 \\
AS7     & 0.903 & 3.2 & 0.999 & 31.6 \\
AS15    & 0.802 & 2.3 & 0.994 & 12.9 \\
GWB7    & 0.837 & 2.5 & 0.997 & 18.3 \\
GWB15   & 0.807 & 2.3 & 0.991 & 10.5
\end{tabular}
\end{ruledtabular}
\end{table}

\begin{figure}
  \includegraphics[angle=270,width=0.95\columnwidth]{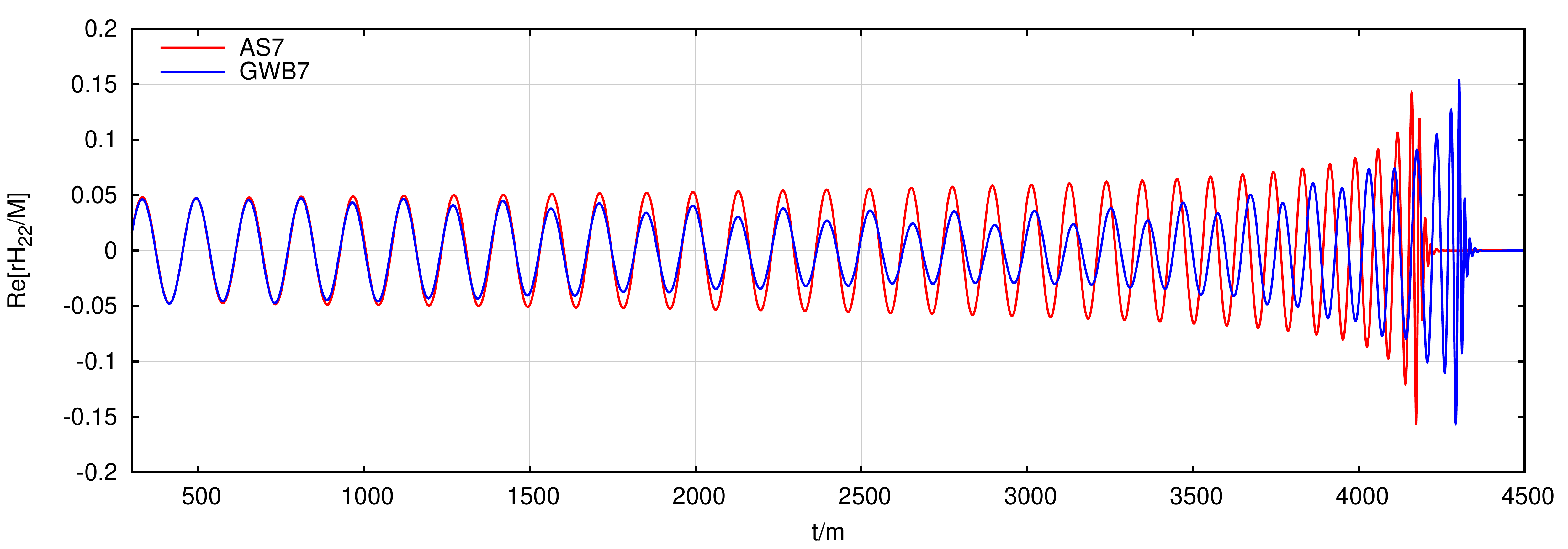}\\
  \includegraphics[angle=270,width=0.95\columnwidth]{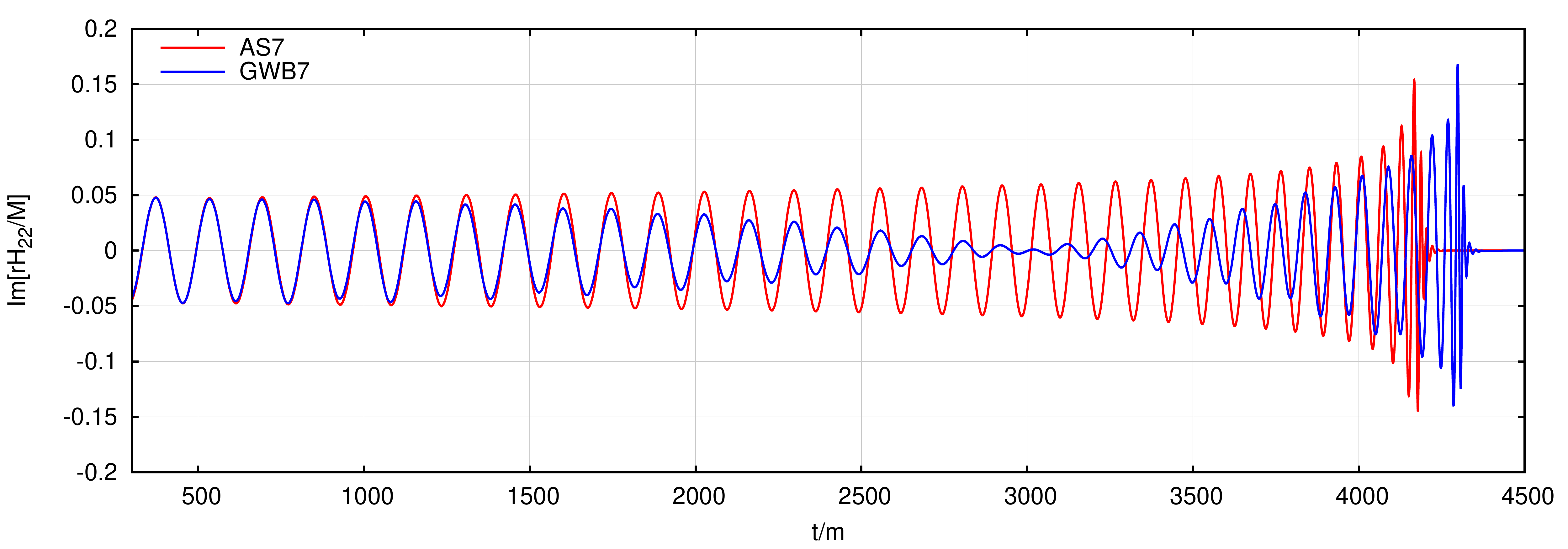}
  \caption{ The real and imaginary parts of the 22 mode of the beaconing and nonprecessing $q=1/7$ case.
In blue (red) is the 22 mode strain waveform of the beaconing (nonprecessing) waveform.
Top panel is the real part, bottom panel is the imaginary part.
Both are extrapolated to infinite observer location.
\label{fig:NRq7}}
\end{figure}

\begin{figure}
  \includegraphics[angle=270,width=0.95\columnwidth]{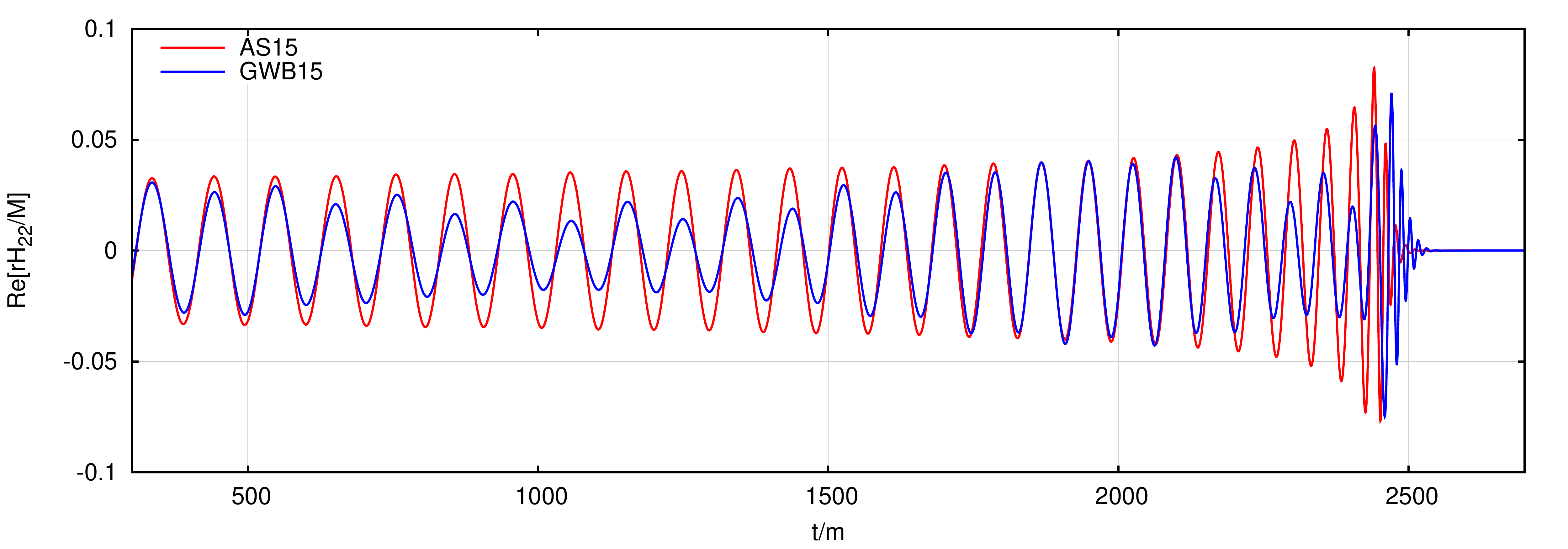}\\
  \includegraphics[angle=270,width=0.95\columnwidth]{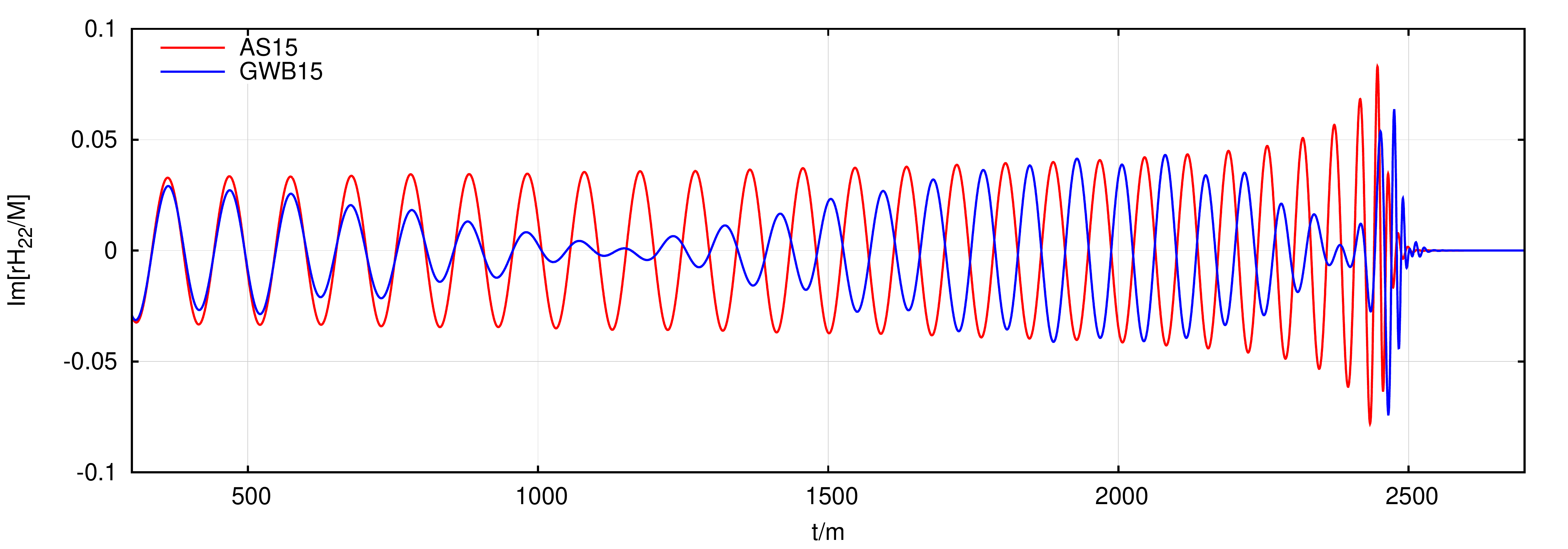}
  \caption{ The real and imaginary parts of the 22 mode of the beaconing and nonprecessing $q=1/15$ case.
In blue (red) is the 22 mode strain waveform of the beaconing (nonprecessing) waveform.
Top panel is the real part, bottom panel is the imaginary part.
Both are extrapolated to infinite observer location.
\label{fig:NRq15}}
\end{figure}

Fig.~\ref{fig:NRq7} shows the NR waveforms for the
22 modes of the strain for the nonprecessing and beaconing $q=1/7$ simulations. 
The waveforms are aligned at the same
initial frequency, and the phase is shifted to agree at the start of the waveform.
Similarly, we show the waveforms for the $q=1/15$ case in Fig.~\ref{fig:NRq15}.
Note in the later, for the imaginary component of the polarization, the clear
change from the ``in-phase'' before the half beaconing period to the exactly
180 degrees ``out-of-phase'' right after,
until the beaconing period is completed.
This shows the expected change from face-on to face-off of the beaconing
binary compared to its nonprecessing counterpart.

To determine if the nonprecessing waveforms are comparable to the beaconing waveforms
along different lines of sight, we ran a series of 390,625 matches over the 
sky angles of both waveforms.  That is, we reconstructed 
$\frac{r}{M} h_{AS}(\theta_{AS},\phi_{AS})$ and $\frac{r}{M} h_{GWB}(\theta_{GWB},\phi_{GWB})$ 
using modes up to $\ell=6$ and varied over $\theta_{AS},\phi_{AS},\theta_{GWB},\phi_{GWB}$
calculating the match between the nonprecessing and beaconing waveforms.  We 
keep the masses constant such that the starting frequency of the waveform
is 30Hz. For the $q=1/7$ case, we find the best match of $0.445$ 
and for $q=1/15$, we find $0.788$. 
This displays the relevance of the black hole spins in the late dynamics
of the binary at the level of gravitational waveforms.

\begin{figure}
  \includegraphics[angle=270,width=0.95\columnwidth]{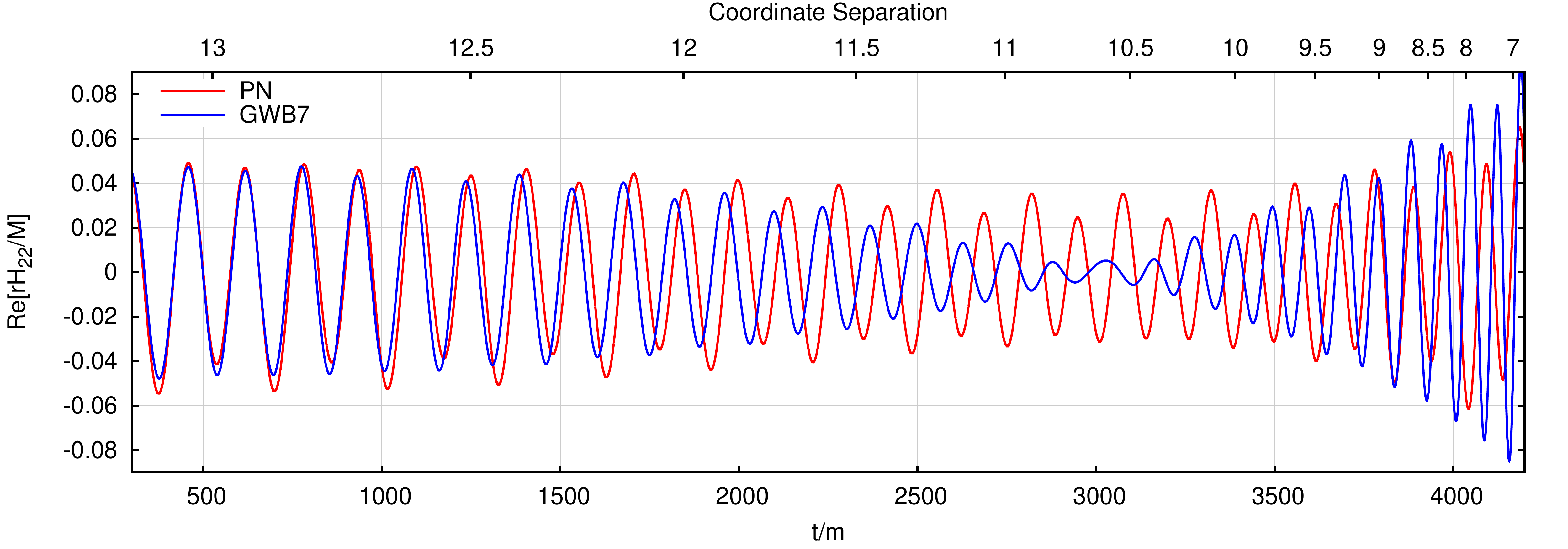}\\
  \includegraphics[angle=270,width=0.95\columnwidth]{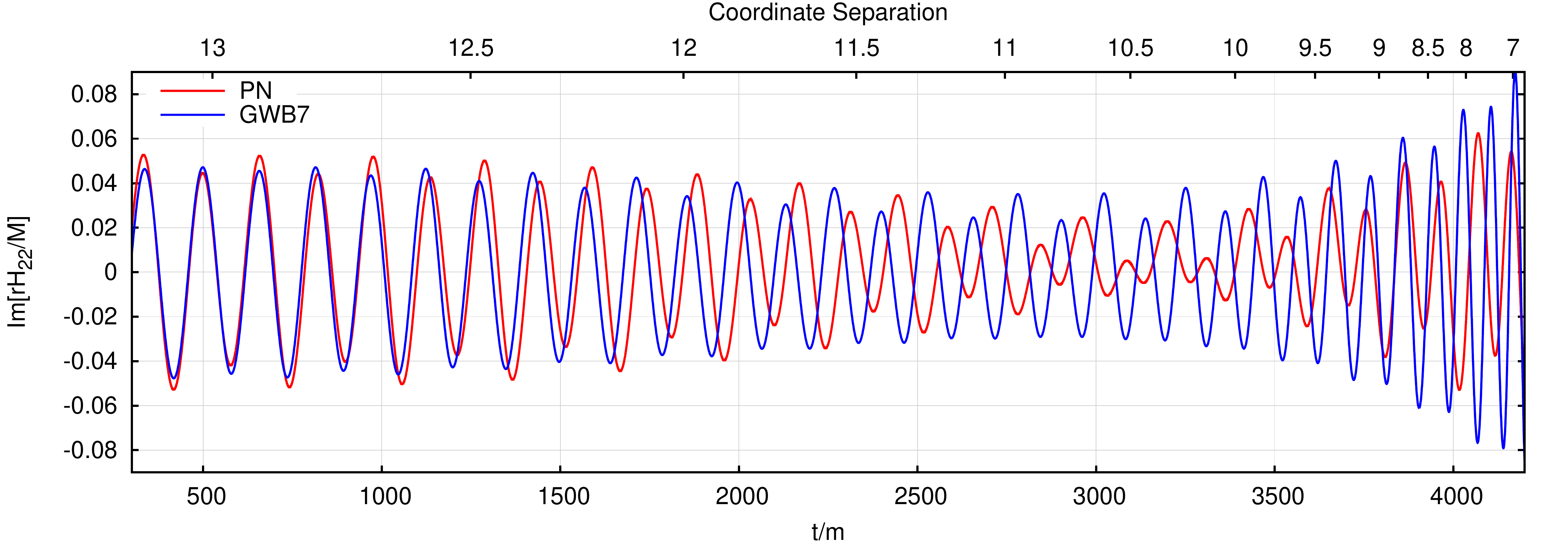}
  \caption{ The real and imaginary parts of the 22 mode of the beaconing $q=1/7$ case.
In blue is the 1.5PN 22 mode strain waveform and 
in red is the NR 22 mode strain waveform extrapolated to infinite observer location.  The
top $x$-axis gives the NR coordinate separation of the waveform and the bottom $x$-axis gives
the time.
\label{fig:GWBq7}}
\end{figure}

\begin{figure}
  \includegraphics[angle=270,width=0.95\columnwidth]{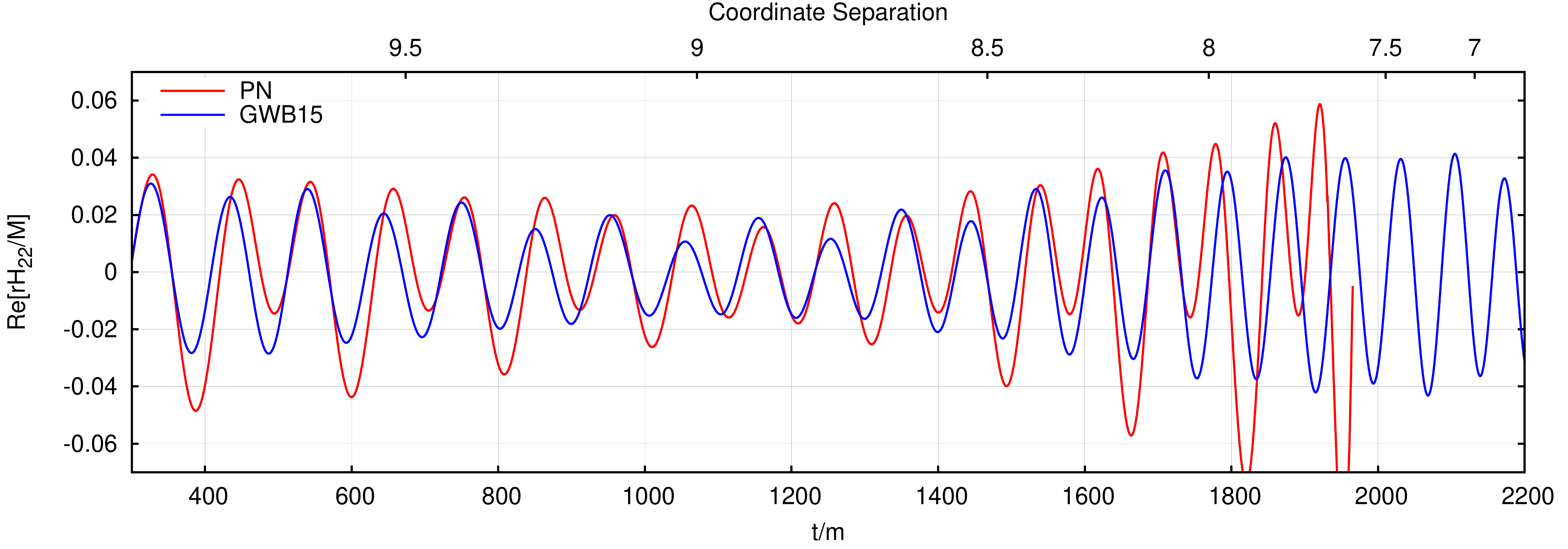}\\
  \includegraphics[angle=270,width=0.95\columnwidth]{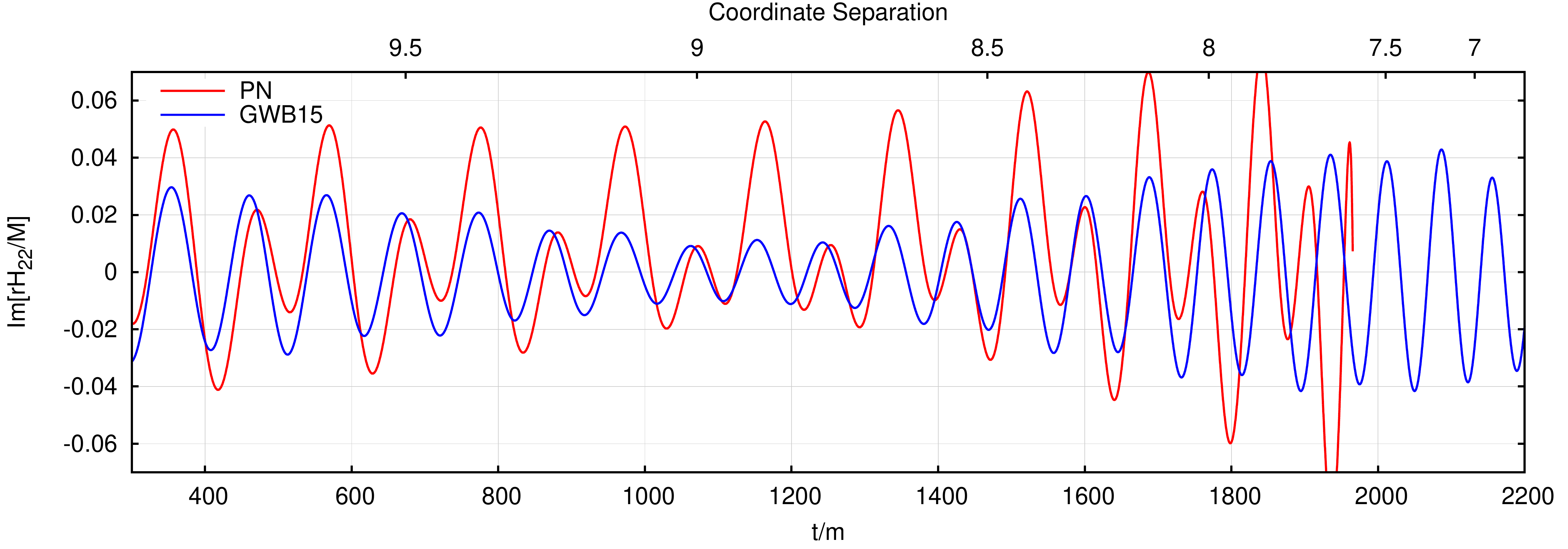}
  \caption{ The real and imaginary parts of the 22 mode of the beaconing $q=1/15$ case.
In blue is the 1.5PN 22 mode strain waveform and
in red is the NR 22 mode strain waveform extrapolated to infinite observer location.  The
top $x$-axis gives the NR coordinate separation of the waveform and the bottom $x$-axis gives
the time.
\label{fig:GWBq15}}
\end{figure}

Figs.~\ref{fig:GWBq7} and~\ref{fig:GWBq15} show the 22 mode NR and PN strain 
waveforms for the $q=1/7$ and $q=1/15$ beaconing cases. The mismatch among them is
apparent and displays the relevance of using accurate NR waveforms for 
the binary's parameter estimation.

\section{Conclusions and Discussion}\label{sec:dis}

We have studied binary black hole configurations that at the time
of the late inspiral are caught in the middle of a transitional
precession. This is a relatively common phase for small mass
ratio ($q<1/4$) {\it retrograde} merging binaries to go through. As shown by
our simulations, schematically depicted in Fig. \ref{fig:config},
the spin of the large hole $\vec{S}$ does not have to be 
particularly fine tuned to almost exactly oppose to the orbital 
angular momentum.
The strong dynamics of the merger notably decreases the magnitude
of $\vec{L}$, eventually matching the spin component along it
(while also accelerating the precession frequency), leading to the formation
of the final remnant black hole before the system completes
the transitional precession (falling into a spin-dominated 
simple precessional state, as predicted in \cite{Apostolatos94}).
On the contrary, we find that $\vec{S}$ and
$\vec{L}$ still roughly precess around a mostly unchanged (in direction)
total angular momentum $\hat{J}$, displaying a total L-flip.

These orbital dynamics have the consequence of generating particular patterns of
gravitational radiation: strong oscillations in the amplitude
of the waveforms and distinct behavior in each polarization.
These special features can be important to identify binary parameters in
this stage when observing gravitational waves. It can also break
current models pseudo degeneracies in the determination of 
the binary's orientation and individual black holes' spins.
This highlights the importance of observing the two
gravitational wave polarizations (with current detectors networks
and in new third generation designs), and,
potentially, could allow tracking the observation of these systems
in a multi gravitational frequency band
\cite{Sesana:2016ljz,Vitale:2016rfr,Wong:2018uwb}.

\begin{widetext}
\begin{figure*}
  \includegraphics[angle=270,width=2\columnwidth]{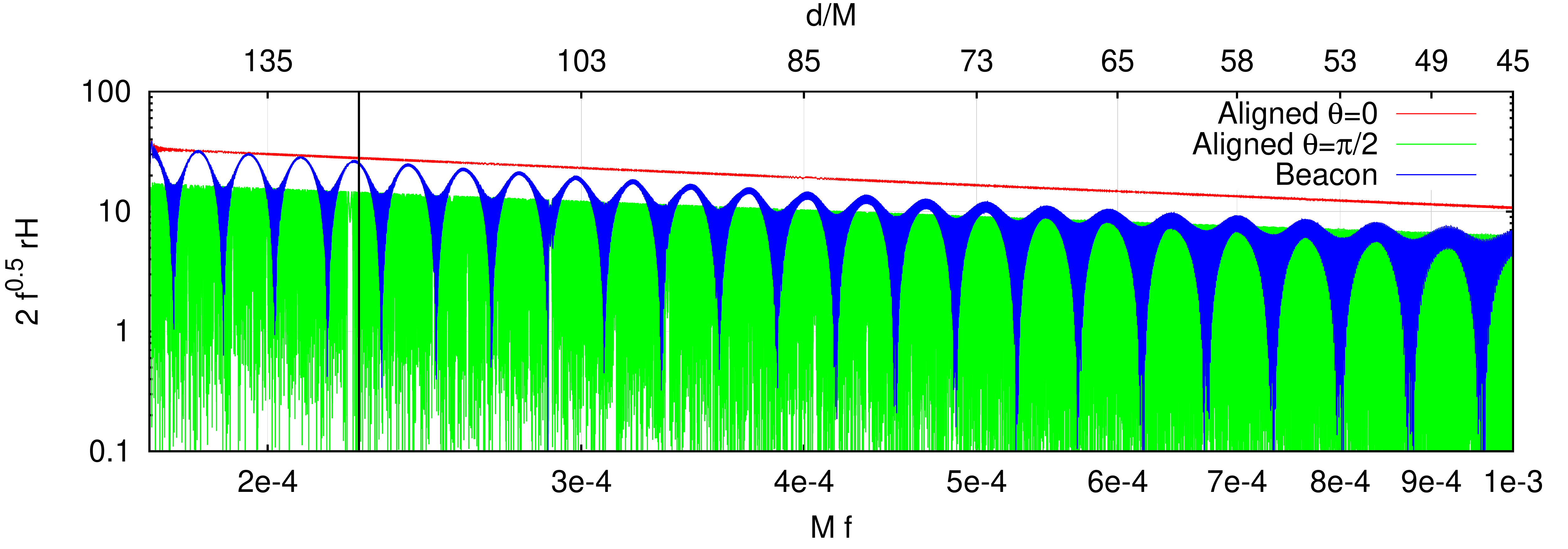}\\
  \caption{The spectral density of the 22 mode of the beaconing $q=1/15$ case 
(evolved with PN from $r=150M$) compared
to the nonprecessing case as seen face-on and face-off.
In blue the strain waveform of the beaconing binary and
in red (green) is the strain waveform of an equivalent spinning but nonprecessing binary as seen from the face-on (face-off).
The top axis gives the separation, r/M, of the nonprecessing binary, and the bottom axis gives the waveform
frequency $Mf$. The black vertical line represents the beaconing frequency before merger. 
\label{fig:d150GWBq15}}
\end{figure*}
\end{widetext}

At the low beaconing frequencies the spectrum of radiation can be estimated via a
PN evolution from an initial separation of $r=150M$ and compared to a nonprecessing
binary with the same initial projected spin along the orbital angular momentum.
Fig.~\ref{fig:d150GWBq15} displays such spectra and shows that the beaconing
amplitude oscillates between the face-on and face-off aligned cases of the nonprecessing binary.
This shows that the beaconing effect gives observers of gravitational waves a much
higher chance of seeing the binary at its peak amplitude.
The strong oscillations in amplitude shown in Figs.
\ref{fig:q7wvf} and \ref{fig:q15wvf}, comparable to the orbital amplitude oscillations, could
lead to the possibility of also detecting systems at lower frequencies,
the {\it beaconing} frequencies.
Let us consider a simple modeling of one of the polarizations of the waveform as
$H(t)=A(t)\cos(\Omega_{L}t)\sin(2\Omega_{orb}t)$, where the amplitude
$A(t)$ varies at the larger inspiral rate, $\Omega_L$ corresponds to the
``beaconing'' frequency scale and provides the envelope of the higher
gravitational wave frequency, that for the leading (2,2) modes is twice
the orbital frequency, $\Omega_L(r_c))$.
Using basic trigonometric identities for the addition of angles
we obtain $H(t)=A(t)\{\sin([2\Omega_{orb}+\Omega_L]t)+\sin([2\Omega_{orb}-\Omega_L]t)\}/2$. Since in general $\Omega_L\ll2\Omega_{orb}$ we have that both
components lie close to $2\Omega_{orb}$ and hence in the higher frequency
band. This is of course a single frequency and single mode analysis and 
the spectrum of gravitational radiation is rather continuous due to the 
merger process and multimode, but this simple model
displays that it is difficult to generate much lower frequency gravitational
waves with significant amplitude. How to detect the beaconing effect at
low frequencies lies outside the scope of this paper, but it is an 
interesting challenge for multiband data analysis.

The results reported in this paper highlight the fact that there is still important
information to extract from observations of gravitational waves.
The relevance of using accurate templates from numerical simulations
was already found for GW170104 in Ref.~\cite{Healy:2017abq}, where a
more robust case for precession is made by directly using
numerical relativity waveforms (see its Fig. 8). 
Not only is the accuracy of the orbital and
spin dynamics important here, but also the inclusion of several (nonleading)
modes (up to $\ell=5$)
to account for effects of the flip of the orbital angular momentum during
the period of observations. 
Note that the phenomenological models, like SEOBNRv3, currently used by LIGO
\cite{Babak:2016tgq} have not been
validated in this range of mass ratios and, as shown in
Refs. \cite{Williamson:2017evr,Healy:2017abq,Blackman:2017pcm},
they differ substantially from NR simulations (e.g., mismatch $\gg1/SNR^2 \sim 10^{-3}$, the maximum allowed to have any chance of consistent parameter inference). Particularly the ad-hoc approach to a precessing merger (not equivalent to the rotation
of a nonprecessing system, as used by the PhenomP model \cite{Hannam:2013oca})
may lead to misevaluations of, for instance, recoil
velocities, here reported in Table \ref{tab:remnant}.

Finally, let us mention that for the last beaconing period to be completely in the 
LIGO band, the total mass of the binaries have to be in a range of masses.
Successive improvements in sensitivity will lead to observations of smaller mass
ratio binaries.
For instance, for the $q=1/15$ binary to start in band (above 20Hz) the total
mass of the system has to be lower than $91M_\odot$ and for the ringdown of the merged
black hole to be in band (below 1000Hz) the total mass has to be higher than
$16M_\odot$. A similar analysis can be done for major mergers sources for
LISA operating in the milliHz band
and the pulsar timing arrays operating in the nanoHz band by scaling up the total mass
of the binaries to the millions and billions of solar masses.

%
%
%

\begin{acknowledgments}
  The authors thank M. Campanelli, R. O'Shaughnessy and
  Y. Zlochower for discussions on this work.
The authors gratefully acknowledge the National Science Foundation (NSF)
for financial support from Grants
No.\ PHY-1607520, No.\ PHY-1707946, No.\ ACI-1550436, No.\ AST-1516150,
No.\ ACI-1516125, No.\ PHY-1726215.
This work used the Extreme Science and Engineering
Discovery Environment (XSEDE) [allocation TG-PHY060027N], which is
supported by NSF grant No. ACI-1548562.
Computational resources were also provided by the NewHorizons,
BlueSky Clusters, and Green Prairies
at the Rochester Institute of Technology, which were
supported by NSF grants No.\ PHY-0722703, No.\ DMS-0820923, No.\
AST-1028087, No.\ PHY-1229173, and No.\ PHY-1726215.
\end{acknowledgments}


\appendix

\section{Merger properties of the nonprecessing and nonspinning simulations}

For the sake of completeness and future reference we provide the remnant properties of the nonspinning and nonprecessing simulations here.

\begin{table*}[!ht]
  \caption{Remnant properties: final mass $M_{f}/M$ and spin $\alpha_{f}$,
    peak radiation $\mathcal{L}^{peak}$ and frequency $M\Omega^{peak}_{22}$,
    and recoil velocity $V_{recoil}$.
The final mass and spin are measured from the apparent horizon, and the
recoil velocity, peak luminosity, frequency, and amplitude are calculated from the gravitational waveforms.
After each column, the percent difference between the measured value and the analytic fit \cite{Healy:2018swt}.
\label{tab:remnantNPNS}}
\begin{ruledtabular}
\begin{tabular}{ccccccccc}
  Case &  NS7 & \%Diff. & NS15 & \%Diff. & AS7 & \%Diff. & AS15 & \%Diff.\\
\hline
$M_{f}/M$                                 & 0.9878   & 0.016\%  &    0.9948 & 0.011\% &   0.9909 & 0.004\% &   0.9956 & 0.005\%\\
$\alpha_{f}^z$                            & 0.3364   & 0.067\%  &    0.1884 & 1.810\% &  -0.0791 & 0.058\% &  -0.0283 & 1.595\%\\
$V_{recoil}[km/s]$                        &     95   & 0.744\%  &        33 & 0.085\% &      131 & 2.753\% &       38 & 1.109\%\\
$\mathcal{L}^{peak}$                      & 1.611e-4 & 0.312\%  &  4.611e-5 & 0.529\% & 1.205e-5 & 4.525\% & 3.911e-5 & 2.541\%\\
$M\Omega^{peak}_{22}$                     & 0.3045   & 0.082\%  &    0.2902 & 0.475\% &   0.2631 & 0.377\% &   0.2665 & 2.168\%\\
$\left(\frac{r}{M}\right)H^{peak}_{22}$   & 0.1611   & 0.404\%  &    0.0853 & 0.642\% &   0.1578 & 0.032\% &   0.0836 & 0.471\%
\end{tabular}
\end{ruledtabular}
\end{table*}

\bibliographystyle{apsrev4-1}
\bibliography{../../../Bibtex/references}

\end{document}